\documentclass[useAMS,usenatbib,aas_macros]{mn2e}

\usepackage{epsfig}
\usepackage{lscape,graphicx,colortbl}
\usepackage{amssymb}
\usepackage{natbib}
\usepackage{times}
\usepackage{aas_macros}

\newcommand{\ifm}[1]{\relax\ifmmode#1\else$\mathsurround=0pt#1$\fi}
\newcommand{\kms}{\ifmmode\,{\rm km}\,{\rm s}^{-1}\else km$\,$s$^{-1}$\fi}

\newcommand{\hmsun}{\,\ifm{h^{-1}}{\rm M_{\odot}}}
\newcommand{\msun}{\rm M_{\odot}}

\newcommand{\dd}{{\rm d}}

\newcommand{\be}{\begin{equation}}
\newcommand{\ee}{\end{equation}}
\newcommand{\bea}{\begin{eqnarray}}
\newcommand{\eea}{\end{eqnarray}}
\newcommand{\z}{\emph{z}}

\newcommand{\fof}{{\scshape fof~}}

\def\ms{m_{\rm star}}
\def\mc{m_{\rm cold}}
\def\mh{m_{\rm hot}}
\def\mg{m_{\rm gas}}
\def\me{m_{\rm ejc}}
\def\mt{m_{\rm total}}
\def\dotms{\dot{m}_{\rm star}}
\def\dotmg{\dot{m}_{\rm gas}}
\def\dotmc{\dot{m}_{\rm cold}}
\def\dotmh{\dot{m}_{\rm hot}}
\def\dotM{\dot{M}_{h}}
\def\fs{f_{\rm s}}
\def\fst{\tilde{f}_{\rm s}}

\def\fc{f_{\rm c}}
\def\ffd{f_{\rm d}}
\def\fa{f_{\rm a}}

\begin{document}

\title[Hydro simulations and semi-analytic models]
      {Hydrodynamical simulations and semi-analytic models of galaxy formation: two sides of the same coin}

\author[E. Neistein et al.]
{Eyal Neistein,\thanks{E-mail:$\;$eyal@mpe.mpg.de}$^1$
Sadegh Khochfar,$^1$ Claudio Dalla Vecchia,$^1$ Joop Schaye$^2$
\\ \\$^1$ Max-Planck-Institute for Extraterrestrial Physics, Giessenbachstrasse 1, 85748 Garching, Germany
\\$^2$ Leiden Observatory, Leiden University, P.O. Box 9513, 2300 RA Leiden, the Netherlands}


\date{}
\pagerange{\pageref{firstpage}--\pageref{lastpage}} \pubyear{2011}
\maketitle

\label{firstpage}


\begin{abstract}
In this work we develop a new method to turn a state-of-the-art
hydrodynamical cosmological simulation of galaxy formation (HYD)
into a simple semi-analytic model (SAM). This is achieved by
summarizing the efficiencies of accretion, cooling, star formation,
and feedback given by the HYD, as functions of the halo mass and redshift.
The SAM then uses these functions to evolve galaxies within
merger-trees that are extracted from the same HYD.
Surprisingly, by turning the HYD into a SAM, we conserve the mass of
individual galaxies, with deviations at the level of 0.1~dex, on an
object-by-object basis, with no significant systematics.
This is true for all redshifts, and for the mass of stars and gas components,
although the agreement reaches 0.2~dex for satellite galaxies at low redshift.
We show that the same level of accuracy is obtained even in case the
SAM uses only one phase of gas within each galaxy. Moreover, we
demonstrate that the formation history of one massive galaxy
provides sufficient information for the SAM to reproduce
the population of galaxies within the entire cosmological box.
The reasons for the small scatter between the HYD and SAM galaxies are:
a) The efficiencies are matched as functions of 
the halo mass and redshift, meaning that the evolution within merger-trees
agrees on average.
b) For a given galaxy, efficiencies fluctuate around the mean value
on time scales of 0.2-2 Gyr.
c) The various mass components of galaxies are obtained by
integrating the efficiencies over time, averaging out these
fluctuations.
We compare the efficiencies found here to standard SAM recipes and
find that they often deviate significantly. For example, here the HYD shows smooth accretion
that is less effective for low mass haloes, and is always composed of hot or dilute
gas; cooling is less effective at high redshift; and star formation changes only
mildly with cosmic time. The method developed here can be applied
in general to any HYD, and can thus serve as a common language for both HYDs and SAMs.
\end{abstract}




\section{Introduction}
\label{sec:intro}

The formation and evolution of galaxies within our Universe is a complicated
process that combines two very different mechanisms. On the one hand, the hierarchical
growth of dark-matter structure drives the aggregation of galaxies, on
time-scales that are proportional to redshift \citep{Press74,Lacey93}.
On the other hand, the baryonic physics determines the interplay between gas
and stars, on time scales that are affected by local processes of
cooling, star formation (SF) and feedback \citep{White78,Dekel86,White91,McKee07}.
The combination of these two disciplines shapes the complex evolution of galaxies
over cosmic time.

Models that take into account the above processes differ in their level of complexity,
and in the typical scales that are being resolved or properly modeled. In general,
a simple distinction can be made between two different approaches,
namely hydrodynamical simulations (hereafter HYDs), and semi-analytic models (SAMs).
HYDs try to follow the evolution of a galaxy, by modelling in great detail the
hydrodynamics and gravitation laws that are in play. These models often use
more than $10^6$ particles to describe one galaxy, and thus allow
its detailed structure to be explored.
However, HYDs are still limited to a finite resolution, which does not allow 
all the processes mentioned above to be followed properly.
Consequently, HYDs rely on `sub-grid' analytical laws, that
describe SF, feedback, and the structure of the inter-stellar medium
(ISM). For a few examples of HYD studies, see
\citet{Katz96,Governato99,Springel03,Scannapieco09,Schaye10,Agertz11}.

A different approach, adopted by SAMs, is to treat each galaxy as one unresolved
object, using integrated properties to describe the mass of stars,
cold gas, hot gas, and the black hole. Since each component of the galaxy
is represented by one number, the dynamics within the galaxy
is not resolved, and one needs to come up with laws for star formation,
cooling, and feedback that are valid on average for the entire
galaxy.\footnote{In more detailed SAMs, that model e.g. the SF rate as a function
of the disk radius \citep{Dutton09,Fu10} one needs to assume an ad-hoc density profile
within the disk.}
Due to their simplicity, SAMs can provide a statistical
sample of galaxies, and can explore a large portion of
their parameter space. For more details, the reader is referred
to some recent SAM studies:
\citet{Monaco07,Somerville08,Benson10,Guo11,Wang11,Khochfar11}.

In the last two decades, HYDs and SAMs have been used as the two major tools
for studying the formation and evolution of galaxies. Detailed comparisons
between the two approaches are thus important both for developing better models,
and for having a common language to interpret different models. Following this
reasoning, various comparisons between the two methods were made to
date. Most of these studies have focused on the processes of
accretion and cooling, finding some agreement at low redshift, and larger
deviations at high-$\z$. For more details, see
\citet{Benson01,Yoshida02,Helly03,Cattaneo07,Viola08,Saro10,Lu11,Hirschmann11}.
In each of the above works, both the SAM and the HYD are adopting a specific
model with a given parametrization. Thus, it is not clear if the discrepancies found
between the HYD and SAM galaxies are due to the limitation of each methodology, or are
just a simple outcome of the specific model chosen.
A few other works have tried to quantify the physics of HYDs without using
a SAM \citep[e.g.][]{Hernquist03,Rasera06,Dave11}. Although such studies
are important for understanding the physics of galaxy formation, it is difficult
to estimate the accuracy of these models for individual objects.

Recently, \citet{Stringer10} have tried a different path to attack
this issue, by trying to tune a SAM according to the physics of a
HYD. These authors managed to modify a SAM based on
\citet{Bower06}, so that it will roughly reproduce the history of
one disk galaxy within a HYD. Since their work was
based on only one galaxy, and since some deviations
between their SAM and the HYD remained, it is still not clear how
well the two methodologies agree.

In this work we would like to take this approach one step further.
We will develop a method to extract the physics of a HYD using the simulation
output, in a way that can be used within a SAM. We use a large cosmological
hydrodynamical simulation, based
on state-of-the-art physical modelling, as developed by
\citet{Schaye10}. Our task is to explore the level of complexity
needed by a SAM in order to follow accurately the formation
histories of galaxies as modeled by the HYD within a large cosmological box.

This paper is organized as follows. In section \ref{sec:methods} we
describe the HYD and the SAM used here, and the method being used to
extract SAM ingredients out of the HYD. These ingredients are presented in
section \ref{sec:hyd_physics}, emphasizing the differences in comparison
to standard SAMs. The galaxies produced by both models are compared in section
\ref{sec:comparison}.
A model with one gas phase is presented in section \ref{sec:1phase},
showing a similar match to the HYD as in the case of the standard SAM.
In section \ref{sec:tests} we further discuss a few additional tests of
the formalism, and try to pin down the reasons for its success.
Lastly, we summarize and discuss the results in section \ref{sec:discuss}.


\section{Methods}
\label{sec:methods}

\subsection{The hydrodynamical simulation (HYD)}

In this work we use a cosmological hydrodynamical simulation (HYD) based on
the OverWhelmingly Large Simulations (OWLS) project \citep{Schaye10}.
This project includes a large set of HYDs
with various different physical ingredients that were studied extensively by
e.g. \citet{Sales10,Wiersma11,vdVoort11,McCarthy11}.
Here we only use one simulation setup, the same as the `reference model' developed by
\citet{Schaye10}. In brief, this model includes radiative
cooling based on \citet{Wiersma09}, following the contributions from 11 different
elements that are released by stellar winds from massive stars, AGB stars
and by supernovae of types Ia and II, as described in \citet{Wiersma09b}.
The SF law is guided by the observed Kennicutt-Schmidt
law, implemented in the form of a pressure law as described in
\citet{Schaye08}. Supernova (SN) feedback
is modeled by injecting SN energy in kinetic form, following \citet{DVecchia08}.
This model includes neither active galactic nuclei (AGN) nor black holes.

We ran a new simulation that is identical to the OWLS reference model with
a box size of 100 $h^{-1}$Mpc, and $2\times512^3$ particles of dark-matter,
gas and stars.
The simulation outputs were saved in 68 snapshots, more than the original run,
and approximately spaced by 200 Myr, from $\z=20$ to $\z=0$.
The dark-matter particle mass equals $4.06\times10^8 \hmsun$,
and baryonic particles have initial mass of $8.66 \times10^7 \hmsun$.
The comoving (Plummer-equivalent) gravitational softening is 7.8 $h^{-1}$kpc
(with a maximum value of 2 $h^{-1}$kpc in
physical units). The underlying cosmological parameters are:
($\Omega_m$, $\sigma_8$, $n_s$, $\Omega_b$, $h$)=
(0.238, 0.74, 0.951, 0.0418, 0.73), consistent with
the WMAP 5-year data \citep{Komatsu09}.

On each output snapshot we have run the \fof algorithm with a linking length of
0.2 \citep{Davis85} to identify haloes with more than 20 dark-matter
particles. The \textsc{subfind} algorithm \citep{Springel01} was then used to identify
subhaloes with more than 20 particles within haloes
(i.e. the minimum subhalo mass ranges between 1$\times10^9$ to 8$\times10^9 \hmsun$, depending on
which particles are included). In our implementation,
\textsc{subfind} uses both dark matter and baryonic particles \citep{Dolag09}.
Since satellite subhaloes within a
dense environment are often being stripped of their dark-matter, subhaloes
occasionally host only star and gas particles.
In addition, fragmentation might happen within haloes,
creating new subhaloes, with only gas and star particles.
We have constructed merger-trees
of subhaloes in the same way as described in \citet{Springel01}.
The trees include information on the subhaloes and their host \fof groups.

\subsection{The semi-analytic model (SAM)}
\label{sec:sam}

Here we describe the semi-analytic model (SAM) used in this work. For
more details on the model, including various specific scenarios for galaxy evolution,
see \citet{Neistein10}. The model follows galaxies inside the complex structure
of merger-trees, and uses simple laws for cooling, SF, accretion, merging, and
feedback. Unlike other SAMs, these laws are simplified to be functions
of only the host subhalo mass and redshift.

\subsubsection{Quiescent evolution}

Galaxies that do not experience merger events are termed to evolve `quiescently'.
Each galaxy is modeled by three phases of baryons,
\begin{eqnarray}
{\rm a \,\, galaxy:} \,\,\,\,\left(  \ms, \, \mc, \, \mh \right)
\, .
\label{eq:m_vec}
\end{eqnarray}
The definitions of $\mc$ and $\mh$ are motivated by the HYD: $\mc$ is the
mass of the cold and dense gas 
that is able to form stars (temperature smaller 10$^5$K, density larger than 0.1 cm$^3$), 
$\mh$ is all
the rest of the gas within the host subhalo, including gas that was previously
inside the subhalo but was later ejected. The exact definitions of the different
gas phases are given in the next section. We note that there exist various different
definitions for cold and hot gas in the literature. Although our definition agrees
with the approach adopted by SAMs, it is different from recent studies based on HYDs, 
as will be discussed below. In addition to our standard model,
we will test various scenarios with differing number of gas phases.

In the following we lay out the basic set of differential equations that describe
the evolution of these phases using a small set of a priori physical assumptions.
These equations have been the basis for the standard paradigm of galaxy formation for over 30 years now
\citep{Rees77,Silk77,White78,White91}.

A fresh supply of gas into the galaxy is provided by smooth accretion
along with the growth of the host dark-matter subhalo. The efficiency of
hot accreted gas is modeled by
\begin{eqnarray}
\left[\dotmh\right]_{\rm accretion} = \left\{ \begin{array}{ll}
\fa\cdot\dotM & \,\,\,\,\,\,\,\,\,\,\,\,\,\,\textrm{if } \dotM>0\\ \;\;\;\
& \;\;\;\;\;\;\;\;\;\;\;\;\;\;\;\;\;\;\;\;\;\;\;\;\;\;\;\;\;\;\;\;\;\;\;  \\
0 & \,\,\,\,\,\,\,\,\,\,\,\,\,\, \textrm{otherwise}
\end{array} \right.
\label{eq:smooth_acc}
\end{eqnarray}
Here $M_h$ is the subhalo mass, defined as the total mass of dark-matter
particles within the subhalo. $\dotM$ is the growth rate of dark-matter
coming from particles that are not included in other subhaloes (not mergers).
Square brackets are used to identify individual processes,
in this case it is the contribution to $\dotmh$ due to accretion.
We allow  $\fa$ to be a function of the halo mass and redshift,
although in standard SAMs \citep[e.g.][]{Croton06} it is a constant that equals the universal
baryonic fraction,\footnote{For low mass haloes,
reionization introduces a filtering mass scale
that gives lower baryon fractions \citep[e.g.][]{Somerville02}.}
$\Omega_b/\Omega_m=0.1756$. 
In general,
a similar component of cold
accretion might exist. However, as will be discussed below, cold accretion is
negligible due to our definition of cold gas, which includes a threshold in density.

Hot gas may radiate and cool according to
\begin{equation}
\left[\dotmc\right]_{\rm cooling} =  -\left[\dotmh\right]_{\rm cooling} = \fc \cdot \mh \,.
\end{equation}
The cooling efficiency, $\fc=\fc(M_h,\z)$, is a function of the host
halo mass $M_h$ and the redshift only, and is written in units of
Gyr$^{-1}$. We assume that the SF rate is proportional to the amount of cold gas,
\begin{equation}
\left[\dotms\right]_{\rm SF} = -\left[\dotmc\right]_{\rm SF} = \fs\cdot \mc \,,
\label{eq:sf}
\end{equation}
where $\fs=\fs(M_h,\z)$ has units of Gyr$^{-1}$.
Gas can be heated due to SN explosions and move from the cold phase into the hot.
In the HYD used here, core collapse SN events follow star formation after a short delay
of 30 Myr. Therefore, the feedback should be proportional to the SF rate,
\begin{eqnarray}
\left[\dotmh\right]_{\rm feedback}    
&=& -\left[\dotmc\right]_{\rm feedback} =  \\ \nonumber
& & \ffd\left[\dotms\right]_{\rm SF} = \ffd \fs \cdot \mc \,.
\end{eqnarray}
Similar to the other ingredients, feedback is modeled by a function of the halo
mass and redshift $\ffd=\ffd(M_h,\z)$.

All the processes above can be united into a set of differential equations,
\begin{eqnarray}
\dotms &=& \fs \cdot \mc \nonumber \\
\label{eq:ode}
\dotmc &=& -\left(\fs+\ffd\fs \right) \cdot \mc + \fc\cdot \mh \\
\dotmh &=&  \ffd\fs  \cdot \mc - \fc\cdot \mh + \fa\cdot \dotM \,.\nonumber
\end{eqnarray}
Each physical process is described by one function ($f_x$), resulting in
a set of linear inhomogeneous differential equations. The hot accretion,
$\fa\cdot \dotM$, is the `source term' that governs the total baryonic mass
within each galaxy. The other three efficiencies ($\fs,\,\fc,\,\ffd$) define the
complex evolution of gas and stars within a galaxy.

\subsubsection{Satellite galaxies}
\label{sec:sam_sat}

In this work we assume that each subhalo includes only one galaxy. 
Since subhaloes might contain only star particles, small subhaloes
inside massive \fof groups can survive longer than in
dark-matter only simulations. We note that although our
SAM uses only the dark-matter mass for each subhalo, the location of the subhalo and its
merging time are affected by the dynamical processes within the HYD, including
contributions from the gas and star particles. Satellite subhaloes are defined
as all subhaloes inside a \fof group except for the central (most massive) subhalo.
Because galaxies and subhaloes have a one-to-one correspondence, we
use the same terminology for central and satellite galaxies.

While satellite galaxies move within their \fof group, they suffer
from mass loss due to tidal stripping. This is modeled by additional terms in the
differential equations above:
\begin{eqnarray}
\dotms &=& \fs \cdot \mc \nonumber \\
\label{eq:ode_sat} \dotmc &=& -\left(\fs+\ffd\fs +\alpha_c\right) \cdot \mc + \fc\cdot \mh \\
\dotmh &=&  \ffd\fs  \cdot \mc - \left(\fc+\alpha_h\right)\cdot \mh + \fa\cdot \dotM \,.\nonumber
\end{eqnarray}
The additional terms including $\alpha_h,\,\alpha_c$ are computed
only for satellite galaxies, and describe the stripping of hot and cold gas respectively.
In general, a similar parameter for stellar stripping can be added, but it is negligible
in the analysis done here. Our model allows for the stripped mass
to be added to the central object, or to be lost to the
inter-galactic medium. For satellite galaxies, all the efficiency
values $\fc$, $\ffd$, $\fs$ are based on the subhalo mass at the last
time the subhalo was central within its \fof group.

\subsubsection{Mergers}

In case a subhalo merges into a more massive one, we merge the
corresponding galaxies as well, and at the same time.
Mergers can trigger SF bursts, with an efficiency that depends on the
mass ratio of the two galaxies:
\begin{equation}
\Delta \ms = 0.56 (m_2/m_1)^{0.7} \times \mc \,,
\label{eq:merger_effic}
\end{equation}
where $m_1\,,m_2$ are the baryonic mass of the central and satellite galaxy
respectively, and $\mc$ is the sum of the cold gas masses of the
two galaxies. This recipe follows the results of hydrodynamical simulations by
\citet{Mihos94} and \citet{Cox08}, and was adopted by various SAMs \citep{Somerville01,
Croton06,Khochfar09,Neistein10}. However, as will be explained below, we do not
find a strong evidence that these bursts are necessary to reproduce
the HYD galaxies, and we therefore do not include bursts in our
final implementation of the model.

\subsection{How to turn a HYD into a SAM}
\label{sec:howto_recipes}

We would like to extract the effective laws that govern the evolution of galaxies
within the HYD. In the language of our SAM, we need to identify the functions
$\fc,\fs,\ffd,\fa$ that summarize the processes of cooling, SF, feedback, and
accretion. For satellite galaxies, we need to determine the constants $\alpha_c$,
$\alpha_h$ that describe the stripping rates of cold and hot gas.
In order to do so we follow each subhalo within the HYD along with
its merger-tree over time, and keep track of all the particle information.
As in the SAM, we assume each subhalo includes exactly one galaxy.

At each snapshot we define three different mass components for each galaxy:
\begin{itemize}
 \item The stellar mass, $\ms$, defined as the total mass of all star particles
 within the subhalo.
 \item The mass of cold gas, $\mc$, is the mass of all particles that are able to
 form stars within the subhalo. According to the SF law being used by the HYD,
 these are all particles with local gas densities larger than 0.1 cm$^{3}$
 and temperatures lower than $10^5$K.
 \item The component of hot gas, $\mh$, includes all gas particles that do not
 belong to $\mc$, as well as all particles that were once within the subhalo, but
 were ejected later. These ejected particles are assigned to the same subhalo
 only if they did not become part of other subhaloes. Note that usually
 in SAMs the ejected gas is treated as a different gas component.
\end{itemize}

We keep track of all particles that belong to subhaloes within the HYD, and
check which of them have changed their phase (i.e. $\mc$, $\ms$, $\mh$) between two
subsequent snapshots, or were accreted/stripped.
For each galaxy $i$ we define all possible transition rates
of the kind $R_{{\rm cold} \rightarrow {\rm star}}^i$,
$R_{{\rm cold} \rightarrow {\rm hot}}^i$,
$R_{{\rm hot} \rightarrow {\rm cold}}^i$, we also checked that other
rates, like $R_{{\rm hot} \rightarrow {\rm star}}^i$ are negligible.
For example, in order to compute the SF rate we use the following sum:
\begin{equation}
R_{{\rm cold} \rightarrow {\rm star}}^i = \frac{1}{\Delta t}\sum_{j} m_j \,.
\end{equation}
Here $m_j$ is the mass of the particle $j$, and the sum goes over all particles
that started as $\mc$ at the beginning of the time-step, and ended as stellar
particles. $\Delta t$ is the time in Gyr between the two snapshots considered.
In order to compute cooling (or heating) rates we use a similar sum, taking into
account all particles that started as hot (cold) particles at the beginning
of the time-step, and ended as cold (hot).\footnote{Multiple transitions of the type
hot$\rightarrow$cold$\rightarrow$hot might exist between two snapshots, but these
are negligible according to the cooling time-scales that will be shown below.
In case multiple transitions exist, not including them will modify the
rates we measure. However, these modifications should not change the mass components of
the SAM galaxies.
This issue reflects the inherent degeneracy of the model equations.}

For the accretion rate,
$R_{\rightarrow {\rm hot}}^i$, we use the sum over all particles that
joined the hot component of the subhalo, and were not identified inside other
subhaloes before. In addition, we
take into account the mass of particles that were exchanged between
subhaloes that belong to different \fof groups. This means that particles
that are stripped into a different \fof group are subtracted from
the accretion rate. On the other hand,
particles that join the central subhalo coming from satellite
subhaloes within the same \fof group are not accounted for in
accretion rates.

Whenever we have a merger event, we first sum up the components of
the progenitor galaxies, and only then compute the rates for the
remnant galaxy. This means that our rates reflect the quiescent
evolution only, and do not include mergers explicitly.
However, mergers might still induce bursts both in the HYD and SAM,
following, e.g., Eq.~\ref{eq:merger_effic}. Mergers can also
affect other processes indirectly, like heating,
cooling or SF within the HYD. This issue will be discussed below.

The efficiencies for each galaxy $i$ are defined by normalizing the rates:
\begin{eqnarray}
\fa^i &=& \frac{R_{\rightarrow {\rm hot}}^i}{\dot{M}_h^i} \,, \\
\label{eq:fc}
\fc^i &=& \frac{R_{{\rm hot} \rightarrow {\rm cold}}^i}{\mh^i} \,, \\
\fs^i &=& \frac{R_{{\rm cold} \rightarrow {\rm stars}}^i}{\mc^i} \,, \\
\ffd^i &=& \frac{R_{{\rm cold} \rightarrow {\rm hot}}^i}
{R_{{\rm cold} \rightarrow {\rm stars}}^i} \,.
\end{eqnarray}
In order to obtain the global efficiency law, for the full cosmological box,
we consider only central subhaloes within their \fof groups.
We then split the sample of subhaloes into bins of different mass
and redshift. For each bin the average efficiency is defined by
averaging the nominator and denominator separately. For example,
\begin{equation}
 \fs(M_h,\z) \equiv \frac{\langle R_{{\rm cold} \rightarrow
 {\rm star}}^i \rangle }{\langle \mc^i \rangle} \,.
\end{equation}
Here averaging is done over all galaxies within the same $M_h$ and
$\z$ bin. Quite arbitrarily, we choose bins of 0.2 dex in $M_h$, and 7
bins in cosmic time, spaced by $\sim2$ Gyr. We have checked that finer
bins do not modify the results of this work.\footnote{The SAM
used here \citep{Neistein10} automatically interpolates the input values of $\fs,\fa,\fc,\ffd$
into a fine grid in halo mass and time.}
The bins in cosmic time are much wider than the
time between two subsequent snapshots. Consequently, the
time average typically includes 10 different snapshots.

\begin{figure}
\centerline{ \hbox{ \epsfig{file=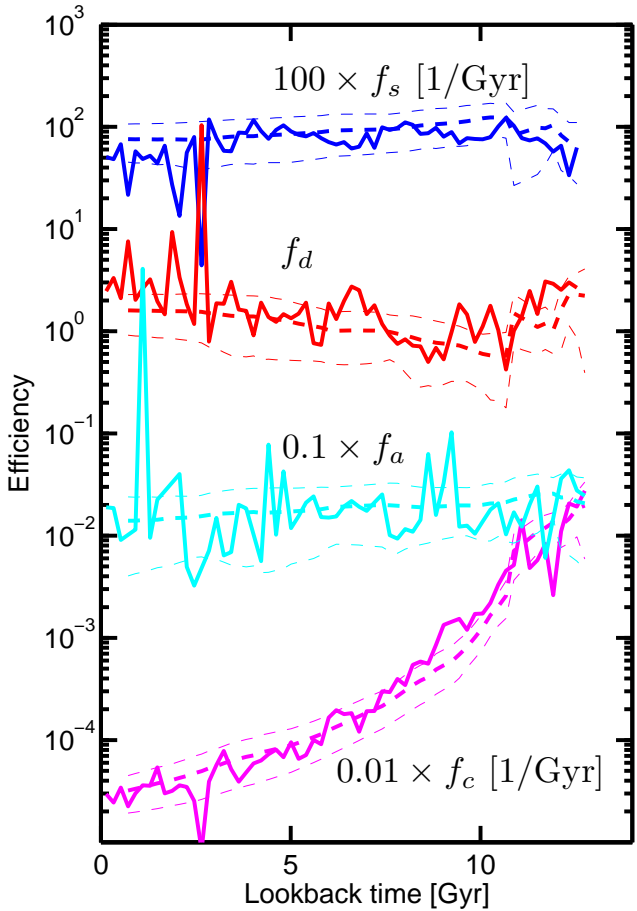,width=9cm} }}
\caption{The efficiencies of cooling, accretion, SF, and feedback following
the main-progenitor history of a specific galaxy within the HYD,
hosted by a subhalo of mass $\sim10^{14}\,\msun$ at $\z=0$.
\emph{Solid lines} show the different efficiencies for this galaxy as measured from the
HYD at all snapshots. The \emph{thick dashed lines} correspond to the
average efficiency for all galaxies within the HYD, including only central galaxies
with the same host subhalo mass and at the same redshift as the galaxy
plotted in solid lines. The \emph{thin dashed
lines} show the standard deviation for the same sample of galaxies above.
The mass components of this galaxy are shown in Fig.~\ref{fig:gal_hist}.}
  \label{fig:gal_rates}
\end{figure}

\begin{figure}
\centerline{ \hbox{ \epsfig{file=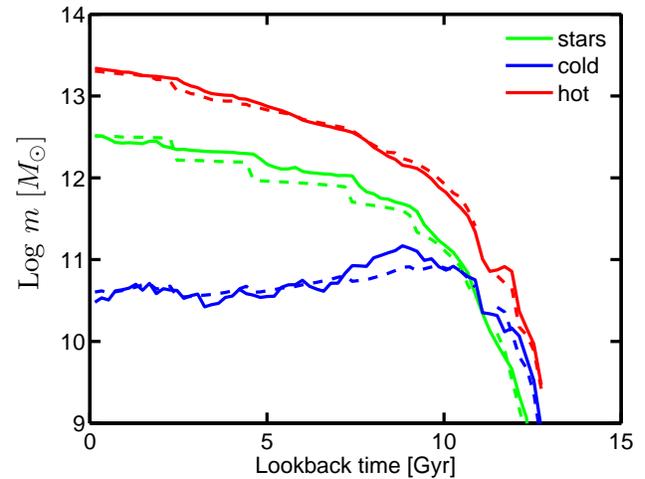,width=9cm} }}
\caption{The baryonic components for the main-progenitor history of
a specific galaxy. \emph{Solid lines} show the mass of stars, cold gas, and hot
gas as measured from the HYD. \emph{Dashed lines} represent the same components
within our SAM, using the same dark-matter subhaloes (see section \ref{sec:comparison}).
The efficiencies of cooling, accretion, SF, and feedback
for the same galaxy are plotted in Fig.~\ref{fig:gal_rates}.}
  \label{fig:gal_hist}
\end{figure}

When computing the accretion efficiency, we use the fact that within the
SAM, negative dark-matter accretion events are treated as zero,
and are not inducing negative gas accretion. To make this approach consistent
with the average value of $\fa$ measured from the HYD, we set all
negative values of $\dotM$ to zero first, only then do we
average $\dotM$ and compute $\fa$:
\begin{equation}
 \fa(M_h,\z) \equiv \frac{\langle R_{ \rightarrow
 {\rm hot}}^i \rangle }{\langle {\rm max}\left(\dot{M}_h,0\right) \rangle} \,.
\end{equation}
This way of averaging guarantees that the total baryonic mass within our
SAM galaxies will agree with the HYD.

We have saved stripping and accretion rates for satellite galaxies, and
recorded the amount of mass flowing into the central subhalo, in comparison
to the total mass being stripped. In general, the stripped mass is best described
by a normalized efficiency, i.e. the ratio $\alpha_h=\Delta\mh/\mh$, as
was defined in Eq.~\ref{eq:ode_sat}. However, we found that both $\alpha_h$ and
$\alpha_c$ are changing as a function of the host subhalo mass and redshift.
In addition, the efficiencies of feedback and cooling for satellite galaxies are somewhat
different than those for central galaxies. This partially depends on the
definitions of the various gas phases, and how each is being stripped.
In this work we have chosen the simplest model possible,
taking into account only constants $\alpha_c$ and $\alpha_h$. Furthermore, we use
the same cooling and feedback efficiencies as for central galaxies. We will show below
that this solution is reasonably accurate for satellite galaxies.
We plan to investigate this issue more closely in a future work.

In the hydrodynamical simulation, star particles can lose some of
their mass due to stellar winds and SN. This mass loss is computed
using a stellar population synthesis model and
is added to the surrounding particles \citep[see][]{Wiersma09b}.
Although this process can be easily modeled within
the SAM, it complicates the interpretation of the results.
This is mainly because the stellar mass loss at a given epoch
is the outcome of the SF history over a few Gyr. Therefore, the rates
measured from the HYD would not be instantaneous, and
might include less scatter with respect to the SAM.
We therefore assume that all particles have a fixed mass, and compute
all rates and efficiencies using this assumption. This assumption
is also being used when comparing the results of the HYD against the SAM.
Consequently, the total baryonic mass within galaxies is sometimes higher
than the universal baryonic fraction. For completeness, we show in
the Appendix all the efficiencies from the HYD when using the proper mass
for each particle, as was used in the simulation.

In Fig.~\ref{fig:gal_rates} we show the different efficiencies for
the main-progenitor history\footnote{The main-progenitor history is defined
by following back in time the most massive progenitor in each merger event.
Note that at high redshift, the subhalo that belongs to the main-progenitor
branch might not be the most massive within its merger tree.}
of one massive galaxy within the HYD, in comparison to the global
averages using all central subhaloes of the same mass and time within the
HYD. It seems that the randomness in the efficiencies of one galaxy
is not too big, and is averaged out over time (except for a few narrow peaks that should not
affect the masses of stars and gas significantly). For example, $\fc$
and $\fa$ show deviations on time-scales of one snapshot (200 Myr),
with no significant trends over larger time-scales. On the other
hand, $\ffd$ and $\fs$ show deviations from the average efficiencies
that are lasting for $\sim2$ Gyr. Overall, the behaviour of one
galaxy seems to be very regular, and does not show significant deviations
larger than the standard deviation (STD)
computed using all the galaxies in the HYD. The total mass in stars,
cold gas, and hot gas for the same galaxy are plotted in Fig.~\ref{fig:gal_hist}.
We will show below that once we use the SAM over the same merger-trees, the
agreement between HYD and SAM is very good, also when comparing
individual objects.


\section{The physics of the HYD}
\label{sec:hyd_physics}

The different efficiencies extracted from the HYD should describe
the various physical processes involved in forming galaxies. As we
will see below, they allow the SAM to accurately reproduce
the population of galaxies in the HYD. This means that we have a
reliable estimate of the net effect that heating, cooling,
accretion, and SF have on galaxies within the HYD.

\subsection{Smooth accretion}

\begin{figure}
\centerline{ \hbox{ \epsfig{file=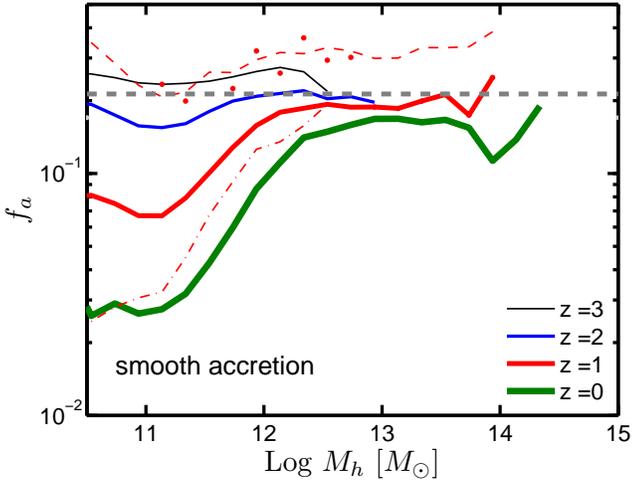,width=9cm} }}
\caption{The smooth accretion rate of baryons as derived from the hydrodynamical
simulation. $\fa$ is defined as the ratio of the smoothed hot accretion, over the
dark-matter smooth accretion, averaged in bins of halo mass
and time. Each \emph{solid line} represents a different redshift bin, all
the other lines are shown only for $\z=1$:
The \emph{dashed line} shows the average plus one standard deviation
in $\fa$; The \emph{dotted-dashed line} shows
the average minus one standard deviation, after averaging out $\fa$ for all
the progenitors within each tree, and at $\sim10$ different snapshots
(all the merger-trees are rooted at $\z=0$);
The \emph{dots} represent the average plus one standard deviation of $\fa$ after averaging
over different progenitors within a tree, but not within different snapshots.
The thick dashed line is the universal baryonic fraction, $\Omega_b/(\Omega_m-\Omega_b)=0.213$.}
  \label{fig:fa}
\end{figure}

The values of the accretion rate, $\fa$, that were extracted from the HYD
are shown in Fig.~\ref{fig:fa}.
We plot only the `hot accretion' component as we do not
detect an accretion of cold gas into galaxies. Although this seems
to be in conflict with various recent studies
\citep[e.g.][]{Keres05,Keres09,Dekel09,vdVoort11} it is a result of the
different definitions of `cold gas' that are being used in the
literature. Here we define cold gas as the gas that is able to form
stars, requiring high densities (larger than 0.1 cm$^3$), and not only low temperatures. This
is a different definition from most other studies based on HYDs,
that often define a gas particle to be cold if it was not previously
heated to the virial temperature of its halo. Here we adopt a more straight-forward
definition of cold gas, based on the SF law. Using our definition,
there is no evidence for `cold accretion' at all redshifts and for all subhalo
masses. This fact is reasonable, because star-forming gas might form stars
before it joins the subhalo, and will therefore be identified as a separate
galaxy.

Unlike in standard SAMs, where $\fa$ is assumed
to be a constant, here $\fa$ shows
a significant dependence on the subhalo mass, decreasing by a factor
of $\sim10$ from subhalo masses of $10^{13}$ to $10^{11}\msun$ at $\z=0$. This is
surprising, considering the fact that all
galaxies used here are the central objects inside their \fof groups.
We have checked this effect further, and tested a model in which
negative gas accretion (stripping) is allowed whenever the dark-matter mass decreases.
Using this new assumption, the accretion efficiencies become much closer
to a constant, with deviations of a factor of $\sim2$,
consistent with \citet{vdVoort11} (the equivalent plot of $\fa$ in
this case is shown in the Appendix, Fig.~\ref{fig:fa_strip}). Our conclusion is that the low
accretion rates shown here for low-mass subhaloes are
compensating for the negative accretion events. Models based on 
\emph{average} dark-matter accretion rates should therefore use the efficiencies
quoted in the Appendix. We have also checked that the total baryonic fraction
within galaxies agrees with the accretion rates given in the Appendix.
In terms of the comparison made here, once we include a mechanism for gas stripping within 
central galaxies, following the dark-matter evolution, we do not get a better agreement
between the HYD and SAM. We therefore adopt the solution of positive
accretion only, as it is more simple to implement.

As was discussed in section \ref{sec:howto_recipes}, the accretion rates shown
here are based on a fixed particle mass,
without taking into account the mass loss due to stellar winds and SN within the
HYD. Consequently, accretion rates can have values larger than the 
equivalent cosmic value, $\Omega_b/(\Omega_m-\Omega_b)$.
In the Appendix we plot all the efficiencies using the proper mass for each particle
within the HYD. This effect changes the overall
normalization of each efficiency slightly, but it does not change the trends
with subhalo mass and time.

\begin{figure}
\centerline{ \hbox{ \epsfig{file=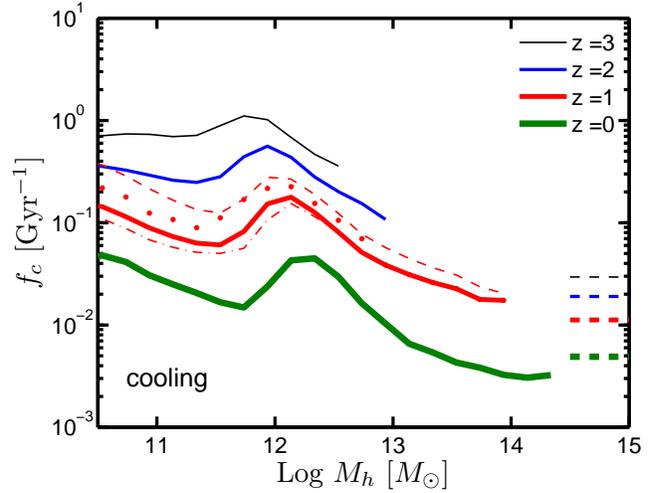,width=9cm} }}
\caption{The cooling efficiencies extracted from the HYD,
and averaged as a function of halo mass and time. $\fc$ is defined
as the ratio of the cooling rate over the mass of hot gas. Plotted lines are using the
same definitions as in Fig.\ref{fig:fa}. \emph{Dashed lines} on the
bottom right are proportional to one over the cosmic time.}
  \label{fig:fc}
\end{figure}

\begin{figure}
\centerline{ \hbox{ \epsfig{file=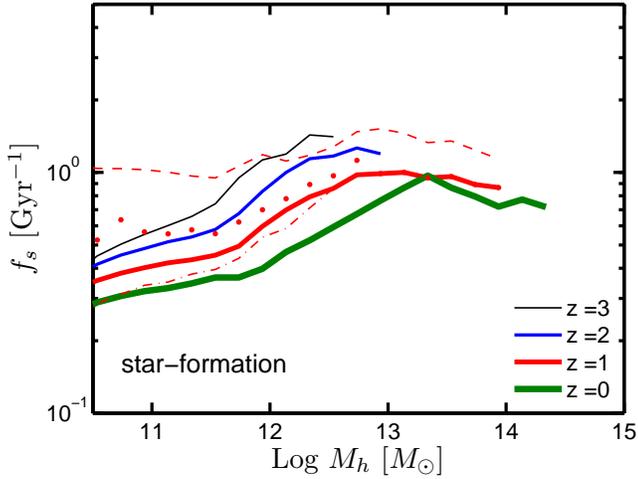,width=9cm} }}
\caption{Star-formation efficiencies within the HYD, defined as the ratio
of the SF rate over the mass of cold gas. For the line definitions
 see Fig.~\ref{fig:fa}.}
  \label{fig:fs}
\end{figure}

\subsection{Cooling}

The next process for which the efficiencies are required is cooling.
In Fig.~\ref{fig:fc} we show average cooling efficiencies for all
the galaxies within the HYD. Here we should keep in mind that the
component of `hot gas' includes gas particles that were ejected out
of the subhalo. Therefore, the cooling efficiencies are normalized
by the sum of both ejected and hot gas, according to Eq.~\ref{eq:fc}.
For a different definition of cooling efficiencies, where hot and ejected phases
are treated separately, we refer the reader to the Appendix.
We note that the dip in the cooling efficiencies seen at a subhalo mass of
$\sim7\times10^{11}\,\msun$ is due to the combination of both hot and
ejected phases.

In general, cooling efficiencies are showing a roughly constant behaviour
as a function of
subhalo masses for subhaloes lower than $\sim 10^{12}\,\msun$, and go down
for more massive subhaloes. This is qualitatively in agreement with semi-analytic
models. However, the dependence of the cooling efficiencies
on cosmic time is stronger than a
simple linear dependence. Since the dynamical time within subhaloes
is proportional to the cosmic time, cooling cannot be modeled only by
the infall time of gas into the centre of haloes.
This might be a result of the cooling process itself, and its
dependence on the hot gas properties \citep[e.g.][]{McCarthy08b,Wiersma09}.

In the terminology
of `cold accretion' mode, where all the accreted gas is assumed to be falling in
narrow streams \citep[e.g.][]{Dekel09}, the process of `cooling' describes the
time it takes the stream to reach the central disk, and become dense enough
to be a part of our definition of $\mc$. A stronger dependence
of cooling on the cosmic time might
mean that accretion through filaments is more relevant at high redshift
\citep{vdVoort11}. A different option is that
trajectories of streams are more radial at high-redshift. This last fact
was already pointed out by \citet{Weinmann11}, and is in agreement
with the orbits of subhaloes within cosmological dark-matter simulations \citep{Wetzel11,
Hopkins10}.

The cooling efficiencies shown here can be compared to
standard SAM algorithms, which are usually following the spirit
of \citet{White91}. This issue was investigated by
various studies in the past \citep{Benson01,Yoshida02,Helly03,Cattaneo07,
DeLucia10,Crain10,Lu11},
mostly claiming some agreement between different SAMs and HYDs, and
some noticeable deviations (especially at high redshift).
For example, \citet{Crain10} showed that the algorithm of \citet{White91}
strongly overpredict cooling rates due to the specific entropy
profiles of gas within haloes. In addition, various SAMs that are
based on \citet{White91} can show significant deviations in cooling
rates due to the detailed implementation of the algorithm
\citep{DeLucia10}.

We find significant
differences when comparing the cooling efficiencies here against the one
used by \citet{DeLucia07} and summarized in \citet{Neistein10}. For example,
at $\z=1$, \citet{DeLucia07} predict cooling efficiency of $\sim1$ Gyr$^{-1}$
at subhalo mass of $10^{11}\,\msun$, roughly a factor of 10 higher than
what is found here.

\begin{figure}
\centerline{ \hbox{ \epsfig{file=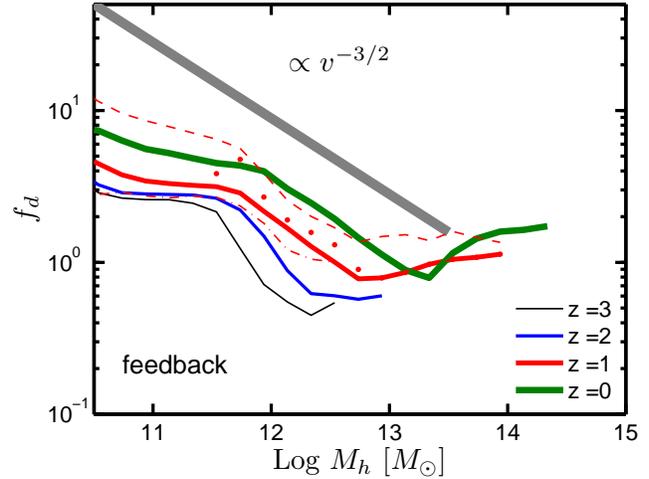,width=9cm} }}
\caption{Feedback efficiencies measured from the HYD, and averaged as a function
of halo mass and time. $\ffd$ is the ratio between the heating rate and the SF
rate. The thick gray line is proportional to the virial
velocity of subhaloes to the power of -3/2, other lines are defined as
in Fig.~\ref{fig:fa}.}
  \label{fig:fd}
\end{figure}

\subsection{Star formation}

Fig.~\ref{fig:fs} shows the SF efficiencies found within the HYD. The time-scales
for converting the cold gas into stars range from $\sim3$ Gyr for low mass
subhaloes at $\z=0$, to $\sim1$ Gyr for massive subhaloes at $\z=3$.
The low-redshift values are roughly consistent with the observational constraints
\citep{Schiminovich10,Saintonge11}. However,
the dependence on redshift found here is much smaller than what is usually assumed in SAMs,
where the conversion efficiency is proportional to the cosmic time
\citep[see, however,][]{Khochfar09}. For example,
\citet{Wang11} showed the SF efficiencies as a function of subhalo mass for
various models, where in standard SAMs the difference between
$\z=3$ and $\z=0$ reaches an order of magnitude.

Interestingly, the SF efficiencies show a double power-law behaviour as a function
of subhalo mass, where the peak efficiency is located at $\sim 10^{12}-10^{13}\,\msun$,
depending on the specific redshift. For high-mass subhaloes the SF is not significantly
reduced. Consequently, the high fraction of passive galaxies within massive subhaloes
is not related to a reduced SF efficiency, but rather to gas consumption, environmental
effects \citep{Khochfar08}, or AGN feedback \citep[e.g.][]{Croton06}.

\subsection{Feedback}

The feedback efficiencies extracted from the HYD are plotted in Fig.~\ref{fig:fd}.
These seem to follow a power-law of the type $v^{-3/2}$ below subhalo mass of $\sim10^{13}\msun$,
where $v$ is the virial velocity of the subhalo. Above this mass, the feedback 
efficiency shows a modest upturn. The feedback efficiency represents gas that is heated
from the cold phase into the hot component, and possibly ejected out of the subhalo.
Within the OWLS reference model, SN feedback is implemented in kinetic form
using a constant wind velocity of 600 km s$^{-1}$. This causes the feedback to become
inefficient for halo masses greater than a few times $10^{11}\,\msun$ \citep{DVecchia08,
Crain09,Schaye10,Haas10}.

SAMs usually assume a power-law efficiency with very different indexes.
For example, \citet{DeLucia07} assumed a constant; \citet{Cole00} have used a power
of -2, following the potential of the host halo; and \citet{Guo11}
assumed a power of -3.5. All these are very different from what is found here.
The feedback efficiency within a model that includes three phases of
gas ($\mc$, $\mh$, and $\me$) is plotted in the Appendix.


\section{Comparing model galaxies}
\label{sec:comparison}

In this section we compare the results of the SAM
with the HYD. The SAM uses only the physical ingredients that were
described above, i.e., $\fa$, $\fc$, $\fs$ and $\ffd$, and was run
using merger-trees extracted from the same HYD. We
specifically use the same values plotted in
Figs.~\ref{fig:fa}--\ref{fig:fd}, using three more bins in cosmic time.
As was explained in section
\ref{sec:howto_recipes}, in order to keep the model simple we do not attempt to model satellite
galaxies with full accuracy.
We set the values of $\alpha_c$ and $\alpha_h$ to 0 and
0.3 respectively, because they provide an effective behaviour which
is similar to that of the HYD. The stripped gas is added to the central galaxy
within the \fof group. Other than that, the SAM has no free
parameters, and no tuning was done.

\begin{figure}
\centerline{ \hbox{ \epsfig{file=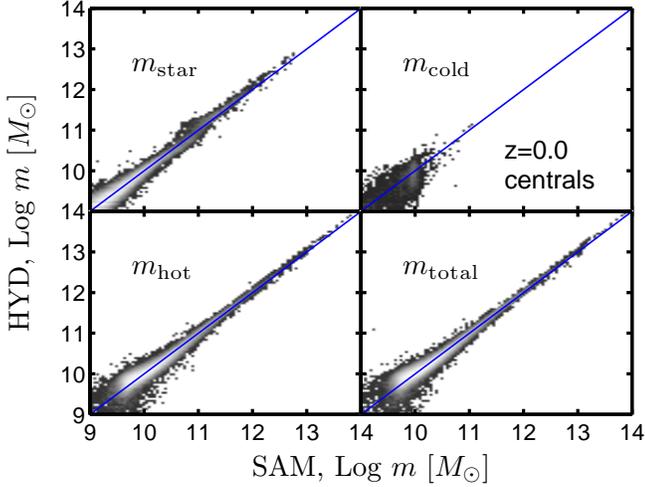,width=9cm} }}
\caption{Comparing the masses of individual galaxies, SAM against the HYD.
Each panel represents a different mass component as labeled, using
only central subhaloes at $\z=0$.
The panels show the two dimensional histogram of the pairs
$(m_{\rm SAM},m_{\rm HYD})$, describing the mass of the same objects in both models.
The pixels are color-coded according to the $\log$ of the number of objects.
The mean difference between HYD and SAM is lower than 0.08 dex for all mass components.
STDs are 0.08 for $\ms$ (except for $\sim10^{11}\,\msun$, for which the STD goes to 0.13 dex).
For $\mh$ and $\mt$ the STD starts at $\sim0.08$ dex for low mass
galaxies and reaches $\sim 0.04$ for
massive galaxies. The STD for $\mc$ is around 0.2 dex.
}
\label{fig:gal_comp_z0}
\end{figure}

The model galaxies from the HYD and our SAM
are compared in Fig.~\ref{fig:gal_comp_z0}. By matching
the same subhaloes from the HYD \& SAM, we are able to show the agreement between
the models on an object-by-object basis. Unlike in previous
studies that showed large deviations between SAM and HYD galaxies \citep[e.g.][]{Hirschmann11}, here the two
models agree quite well. For central galaxies the STD of differences is
less than 0.1 dex, for all redshifts and for the various galaxy components
(except for $\mc$, which usually includes only a few tens of particles within a galaxy).
The total mass in baryons, $\mt\equiv \ms+\mc+\mh$, is shown as a probe of
the accuracy of the accretion rates, $\fa$.
Note that the mass of gas particles within the simulation is $8.64 \times10^7 \,
\hmsun$, so masses below $10^{10}\,\msun$ include
less than 100 particles and suffer from various numerical artifacts, also
within the HYD.

We found that deviations between $\mt$ in the SAM and the HYD correlate
strongly with deviations in $\mh$, and are the reason for most of the scatter
found in $\mh$. This is a consequence of the fact that most of the baryonic
mass is located in $\mh$. A similar (but weaker) correlation exists between deviations in $\mt$ and
$\ms$. Even though the stellar mass is affected from various additional processes that
seem to be more complicated than accretion, the deviations in $\fa$ between the HYD and the SAM
still affect $\ms$. This fact can also
be seen in Figs.~\ref{fig:fa}-\ref{fig:fd}. In these plots we show the STD of
each efficiency at $\z=1$, after averaging out all progenitors of the same
galaxy at $\z=0$ (dotted-dashed lines). It is evident that the scatter in $\fa$
between different galaxies is significantly higher than the scatter in other efficiencies. This might be
a result of different merger-histories for subhaloes that live inside different
environment densities \citep[the assembly bias effect,][]{Gao05}.

The small scatter in $\ms$ between the HYD and the SAM (0.08 dex)
is interesting in view of the larger scatter in $\mc$ (0.2 dex).
Although the masses of $\mc$ are usually below $10^{10}\,\msun$, and are therefore
not numerically reliable, these masses are responsible for making stars, and somehow
produce a small scatter in $\ms$. This issue will be discussed in section \ref{sec:why_agree}.
A different contribution to the scatter in $\mc$ comes from mergers, and will be
discussed in section \ref{sec:mergers}.

\begin{figure}
\centerline{ \hbox{ \epsfig{file=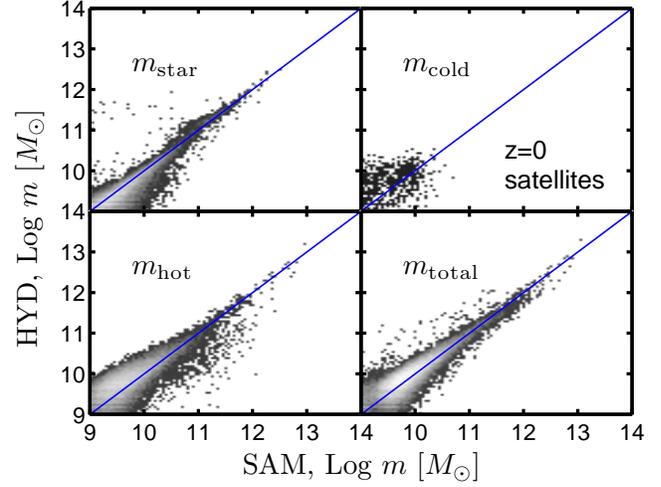,width=9cm} }}
\caption{Comparing the mass of individual galaxies, SAM against the HYD.
Data was derived in the same way as in Fig.~\ref{fig:gal_comp_z0},
but for the population of satellite
galaxies at $\z=0$. Mean (STD) differences between SAM and HYD reach 0.1 (0.2) dex for
both $\ms$ and $\mt$, and 0.3 (0.5) dex for $\mh$.}
\label{fig:gal_comp_z0_sat}
\end{figure}

\begin{figure}
\centerline{ \hbox{ \epsfig{file=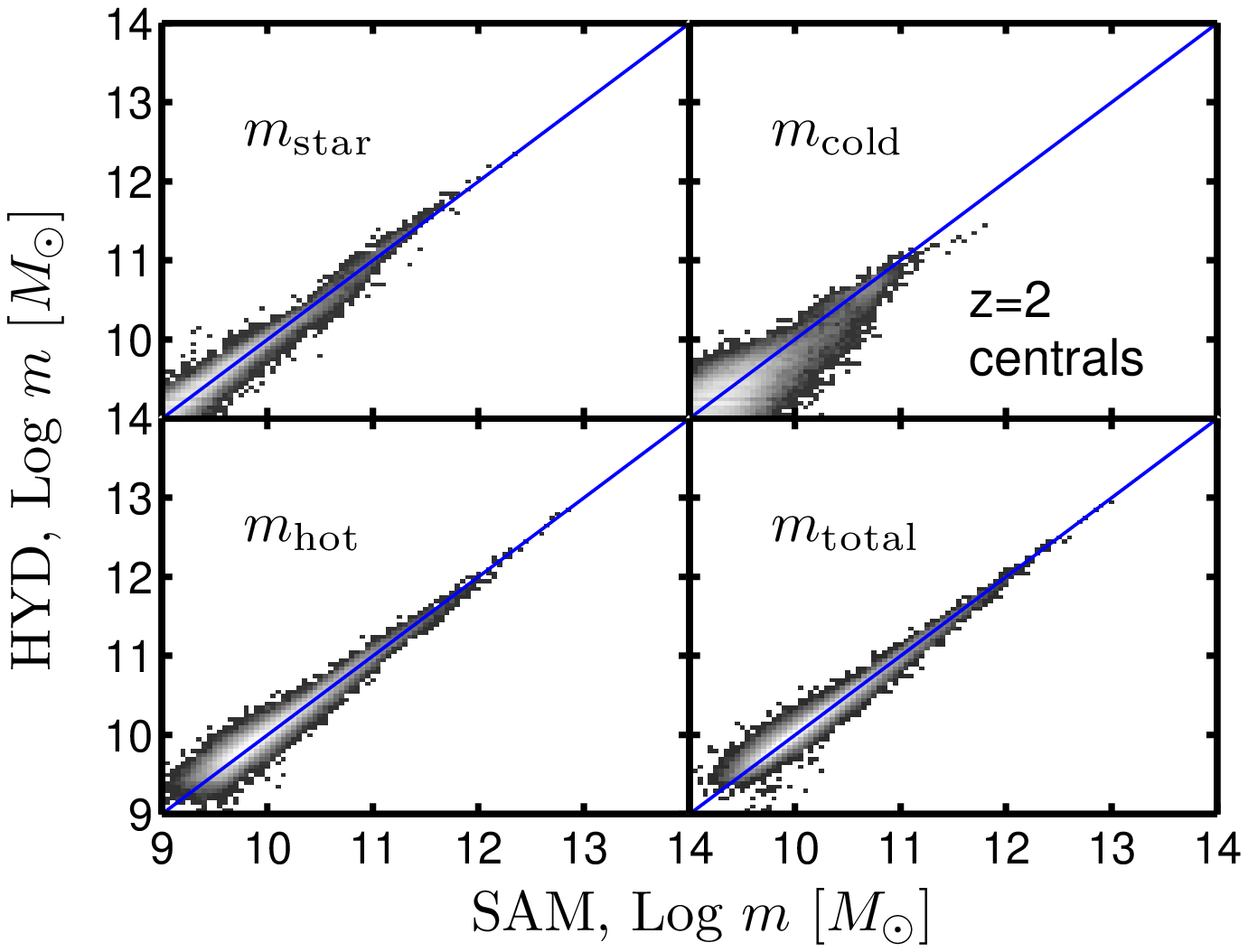,width=9cm} }}
\centerline{ \hbox{ \epsfig{file=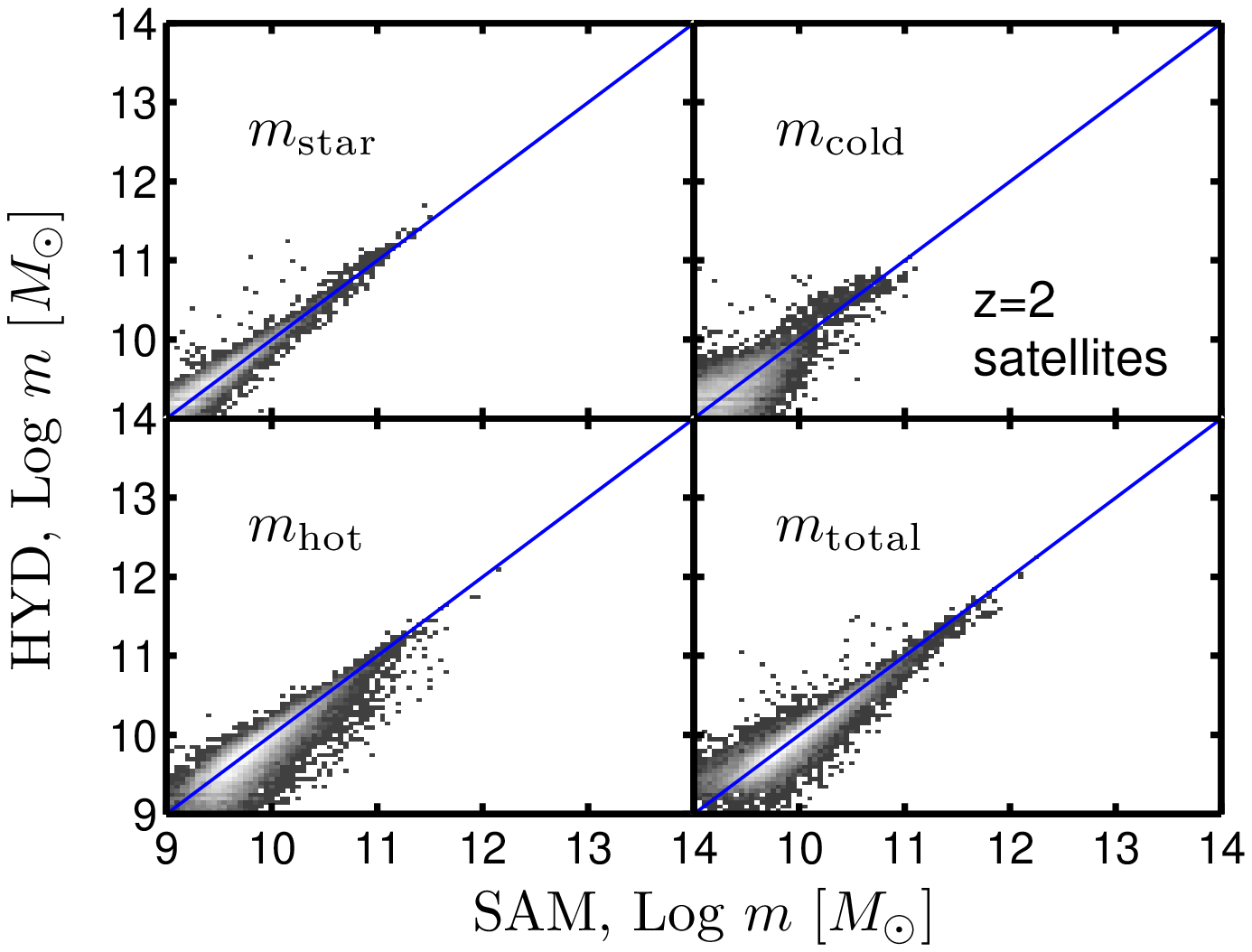,width=9cm} }}
\caption{Comparing the mass of individual galaxies, SAM against the HYD.
Histograms are the same as in Fig.~\ref{fig:gal_comp_z0}, but for the
population of central and satellite galaxies at $\z=2$. For central galaxies,
the mean and STD of differences between HYD \& SAM are similar to the numbers
quoted in Fig.~\ref{fig:gal_comp_z0}. For satellite galaxies the agreement is
roughly two times better than in Fig.~\ref{fig:gal_comp_z0_sat} (in dex units).}
\label{fig:gal_comp_z2}
\end{figure}

\begin{figure}
\centerline{ \hbox{ \epsfig{file=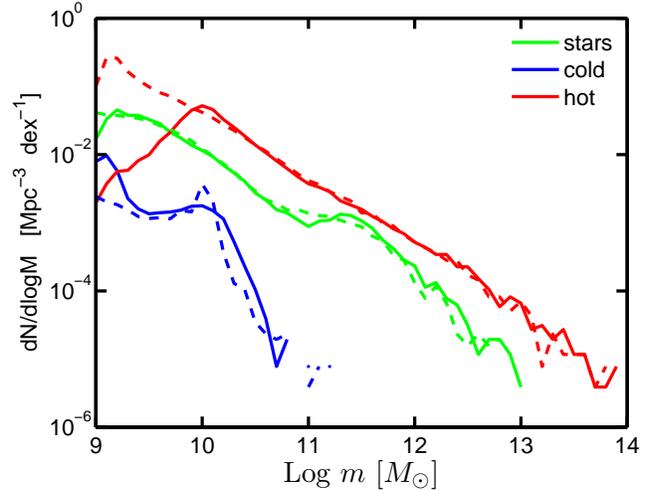,width=9cm} }}
\caption{The mass functions of $\ms$, $\mc$, and $\mh$ in the HYD (solid lines)
and in the SAM (dashed lines). All galaxies at $\z=0$ are selected.}
\label{fig:mass_funs}
\end{figure}

We have explored the larger scatter in $\ms$ at values of $10^{11}$ by
running a different HYD, with a different feedback model (Dalla Vecchia \& Schaye, 
in preparation), and by increasing the
resolution of the efficiency bins (both in mass and time). The high-resolution
efficiencies were not making any noticeable change, but the HYD with a different
feedback model results in a significant smaller scatter at $\ms=10^{11}\,\msun$.
It might be that the kinetic feedback prescription used by the HYD affects the
hydrodynamical state of the gas in a way that is different than other cooling
and heating channels. These changes might complicate the simple distinction done
here between cold and hot gas.
In addition, it might be that the transition between effective and ineffective
feedback is sensitive to other properties of the subhaloes other than the subhalo mass.

A comparison for satellite galaxies at $\z=0$ is shown in Fig.~\ref{fig:gal_comp_z0_sat}.
Here the deviations between the SAM and the HYD are larger than for central galaxies,
reaching 0.2 dex for $\ms$. We have tried a model in which stripping of satellite
galaxies follows the stripping of dark-matter, according to \citet{Weinmann10}.
This model did not improve the match between the HYD and the SAM, probably because
satellite galaxies experience on average different efficiencies of cooling and feedback,
as was discussed in section \ref{sec:howto_recipes}. This can be seen in
Fig.~\ref{fig:gal_comp_z0_sat}, where $\mt$ for satellite galaxies behaves better
than $\ms$ and $\mh$, hinting that the total amount of stripping is modeled properly.
The physics of satellite galaxies
is complicated, and deserves more attention than we give it in this work.

\begin{figure}
\centerline{ \hbox{ \epsfig{file=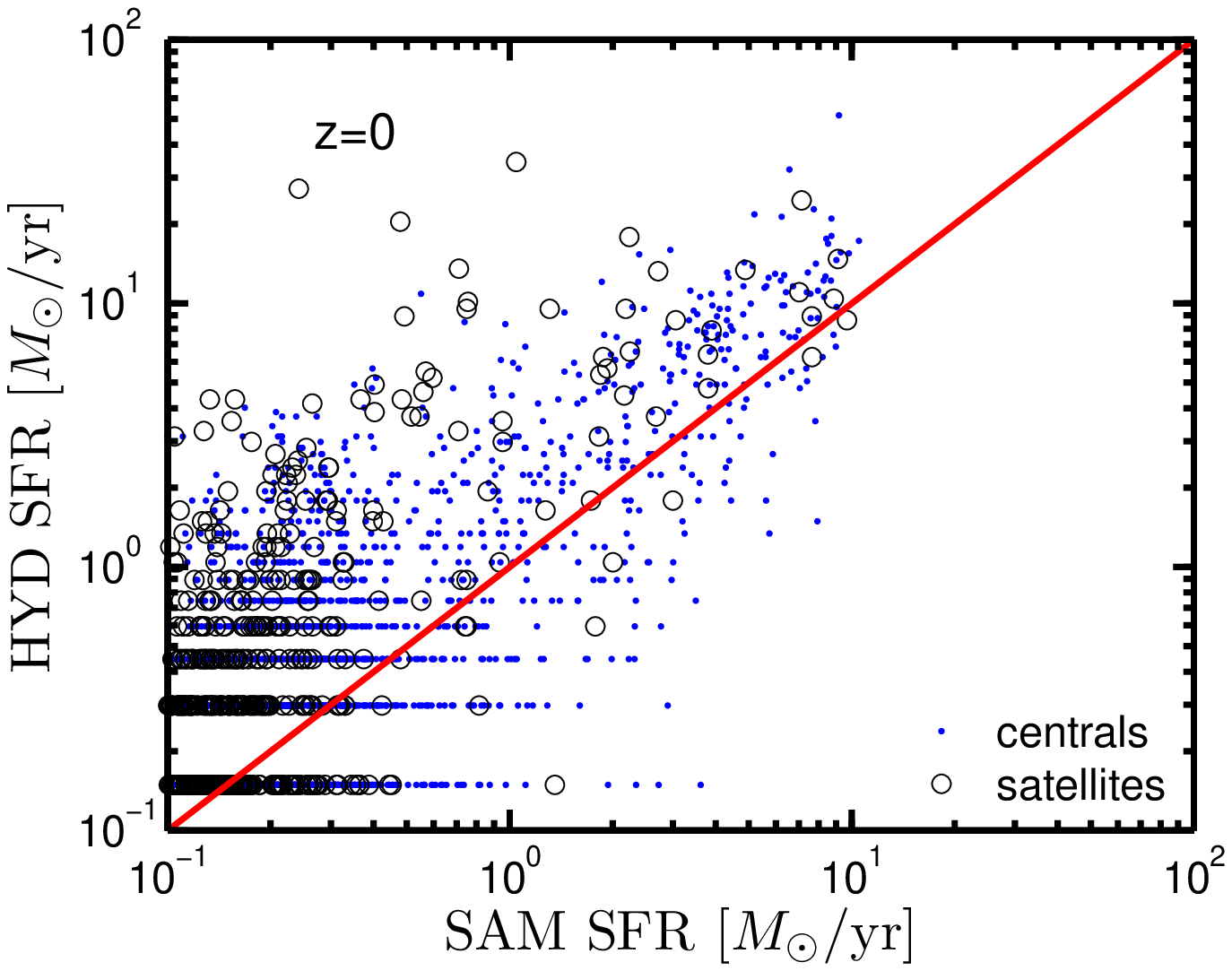,width=9cm} }}
\centerline{ \hbox{ \epsfig{file=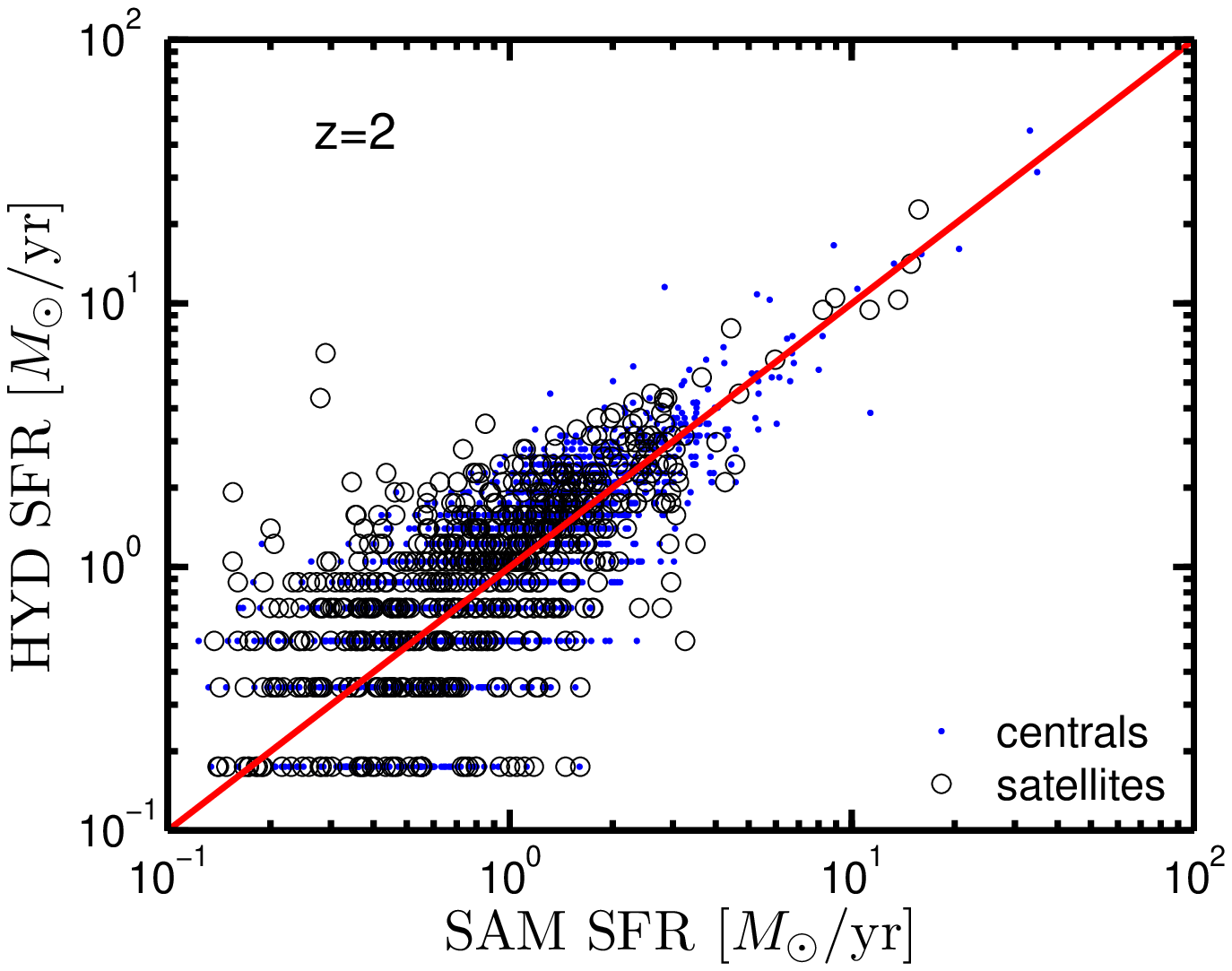,width=9cm} }}
\caption{Comparison of SF rates for individual galaxies within the models,
SAM against the HYD.
Central and satellite subhaloes are plotted in dots and circles respectively.
The solid line shows the values where both SF rates agree. Top and bottom panels show
results for $\z=0$ and $\z=2$, respectively.}
\label{fig:comp_sfr}
\end{figure}

In Fig.~\ref{fig:gal_comp_z2} we compare galaxies at $\z=2$, finding similar
trends to $\z=0$ as discussed above. The mass of cold gas is much higher at this redshift,
so the agreement and deviations are clearer. For satellite galaxies the agreement
is much better than at $\z=0$, probably because these galaxies had much less time
to evolve within their group, and deviations have not accumulated yet.

The mass functions for $\ms$, $\mc$, and $\mh$ using both the HYD and the SAM are shown in
Fig.~\ref{fig:mass_funs}. Overall, the agreement between the two models is very
good for all mass components. This is expected, as the
agreement for individual objects is good. For $\mh<10^{10}\,\msun$ the sensitivity of the HYD
to resolution effects seems to be high. This probably hints to the dependence of the
cooling mechanisms on resolution. Fig.~\ref{fig:mass_funs} includes
all the galaxies within each model. We note that the other figures in
this section only show galaxies that exist \emph{both} in the SAM
and in the HYD. Due to our definitions, galaxies that just emerged
within the HYD (and do not have any progenitors) are not included in the
SAM. On the other hand, the SAM keeps a small population of galaxies
that do not have descendant subhaloes. These two populations are
quite small and do not affect the mass functions.

The SF rates for individual objects are compared in Fig.~\ref{fig:comp_sfr}.
Unlike the integrated properties shown above, the SF rates show
stronger deviations between the two models, with a STD of $\sim0.5$ dex.
The scatter in $\mc$ is about 0.2~dex, meaning that the deviations
in the SF rate are dominated by variations in the SF efficiency.
In addition, most of the population of galaxies at low redshift form stars at a low
rate, $\sim1\,\,\msun$ yr$^{-1}$. This rate corresponds to just a few gas
particles within a snapshot, increasing the scatter between HYDs
and  SAMs. As can also be seen in Fig.~\ref{fig:fs}, the scatter in $\fs$ between
all galaxies is large, reaching a factor of 3 for low mass subhaloes.
However, when $\fs$ is averaged over all the progenitors within a tree,
and over a few snapshots, the scatter goes down dramatically. Apparently,
the deviations in SF rates result in much smaller deviations for stellar masses.
We will examine this issue in section \ref{sec:why_agree}.

\subsection{How important are merger-induced bursts?}
\label{sec:mergers}

Our SAM does not include any merger-induced processes, like SF bursts, 
heating or cooling.
We have tested the contribution of mergers to the models in various ways.
First, we have computed the average baryonic efficiencies after excluding galaxies
that had a major merger event in the last 0.5 Gyr, or that have a major satellite
galaxy at a distance smaller than 0.5 Mpc (`major mergers', and `major satellite'
are defined to have a mass ratio larger than 0.2). Applying this selection criterion
does not change the results of the efficiencies in a noticeable way.
The agreement between the SAM and HYD galaxies does not change either. There is
only a minor change in the average agreement in $\mt$. However,
it might be that the \emph{number} of merger events is small, and
cannot affect the full population of galaxies within a cosmological box.

A second test we carried out is to run our SAM with an additional recipe for
merger-induced SF bursts. We have used the standard recipe
given in Eq.~\ref{eq:merger_effic} above.
However, in terms of the comparison made here, this recipe
does not change the agreement between the HYD and the SAM.

Since the effect of mergers might accumulate with time, we would like to
define a quantity that is related to the number of mergers a galaxy had in its
past. Consider a galaxy at $\z=0$, and define the mass in stars,
$m_i$ ($i>1$), that was accreted from each satellite galaxy $i$ within
the merger tree.
All the stars that were formed within the main-progenitor branch are
termed $m_0$. Using $\ms$, the stellar mass of the galaxy today, we define:
\begin{equation}
\sigma_m = \sqrt{\sum_{i\geq0} \left( \frac{m_i}{\ms} \right) ^2} \,\,\,.
\end{equation}

If the stellar mass of a galaxy is built from
$N_s$ equal values of $m_i$, then $\sigma_m$ would equal $1/\sqrt{N_s}$. Low values of
$\sigma_m$ indicate on many merger events, while high values (close to unity)
indicate that the galaxy is built from one branch only. The meaning
of using $\sigma_m$ can be related to the comparison we make between
the HYD and the SAM. Assume that each progenitor galaxy within the SAM includes some random,
normally distributed error in stellar mass (with respect to the HYD), that is proportional to its mass, $m_i$.
In this case the relative error in the sum of all masses (i.e. the error in the SAM prediction for $\ms$)
will equal $\sigma_m$.

We have computed $\sigma_m$ for each galaxy within our SAM, and
measured the correlation between $\sigma_m$ and the deviations
between the SAM and HYD galaxies. We found that the mass within the SAM
galaxies is higher than within the HYD for galaxies with more mergers
(galaxies with lower $\sigma_m$). This effect is true for all mass
components ($\ms$, $\mc$, and $\mh$), but it is strongest for $\mc$.
This means that galaxies within
the HYD lose some of their mass in each merger event, or that satellite
galaxies lose some of the stripped mass to the inter-galactic medium.
This effect is not modeled by our current SAM, but it should be
straight forward to add it, once the treatment of
satellite galaxies is more accurate.

To conclude, we do not find any significant evidence for
merger-induced SF bursts within the HYD used here. This seems to be in
conflict with previous simulations of galaxy mergers
\citep[e.g.][]{Mihos94,Cox08}. However, as was suggested by \citet{Moster11},
it might be that the presence of hot gas in subhaloes
regulates the efficiency of bursts within our simulation.
It might also be that the mass gained in bursts does not contribute
much to the total stellar mass within galaxies \citep{Khochfar06},
and that mergers are too rare \citep{Lotz08a,Hopkins10}. On the
other hand, we do find significant evidence for mass loss within
mergers, as was pointed out by previous studies \citep{Monaco06,
Purcell07, Conroy07, Yang09a}.


\section{The one-phase model}
\label{sec:1phase}

The analysis above was based on a SAM with two different gas phases within a 
galaxy, $\mc$ and $\mh$. As an alternative, this section describes a model with 
only one phase of gas, allowing us to study the uniqueness of the SAM equations.
The one-phase model has been explored by other studies in the past. In one of the earliest
SAM works, \citet{Cole91} used modelling of cold gas in haloes to predict the
galaxy luminosity function. Although SAMs are usually based on three phases
of gas in galaxies, the one-phase model was recently explored  by
\citet{Bouche10,Krumholz11,Khochfar11a,Dave11}. In what follows, we will try to
emphasize the points of similarity and difference with respect to these previous works.

\subsection{A SAM with one phase of gas}

\label{sec:1phase_eq}

For each galaxy we define
\begin{equation}
\mg = \mh + \mc\,.
\end{equation}
Thus, $\mg$ includes all gas particles within the subhalo, as well as
gas particles that were ejected from the halo. In this case,
the equations that govern galaxy evolution are:
\begin{eqnarray}
\dotms &=& \fst \cdot \mg \nonumber \\
\dotmg &=& -\fst \cdot \mg + \fa\cdot \dotM \,.
\label{eq:ode_1phase}
\end{eqnarray}
Within the SAM, these equations assume that $\fst$ and $\fa$ are
independent of the galaxy components $\ms$ and $\mg$. However, if
we compare this model to the standard model with the two phases above (Eq.~\ref{eq:ode}),
we get that $\fst=\fs\mc/\mg$. Consequently, if the standard model is accurate,
the one-phase model cannot be treated as a set of ordinary differential
equations as we assume below.

The one phase model is useful because it is simple, and in case
$\fst$ does not depend on $\mg$, it can be integrated to follow the evolution
of one object, with no mergers. First, we define the integral,
\begin{equation}
P(t_0,t_1) = \exp \left[ -\int_{t_0}^{t_1} \fst \dd t \right] \,,
\label{eq:P}
\end{equation}
where the values of $\fst$ within the integrand are computed for
the specific mass history of the subhalo. Once $P$ is computed,
the solution of the set of equations can then be written as
\begin{eqnarray}
\label{eq:mg_solution}
\lefteqn{ \mg(t) = \mg(t_0) \, P(t_0,t) + }  \\ & & \nonumber \;\;\;\;
P(t_0,t) \,\int_{t_0}^{t} \frac{\fa(t_1) \dotM(t_1)}{P(t_0,t_1)} \dd t_1
\end{eqnarray}
It should be emphasized that this solution is valid for one branch only, and cannot
be expanded easily to include all the progenitors within a merger-tree. This is because
both $\fa$ and $\fst$ depend on the mass of each progenitor.

A model with one gas phase can be written in various ways. For
example, we could have a model that is based only on the cold gas
within a galaxy, as was done by \citet{Bouche10,Khochfar11a} and \citet{Dave11}.
It is evident from Eqs.~\ref{eq:ode} that a straightforward
way to do it is to assume that the
hot gas component equals the \emph{total} baryonic mass, neglecting
the 3rd equation, and keeping the two first equations almost
unchanged. It should be noted that we then get a set of equations
that is different from the above studies. Here the
source term is not $\fa \dotM$, but $\fc \tilde{f}_{\rm a} M_h$, where
$\tilde{f}_{\rm a}$ represent the fraction of hot gas within the subhalo.
Since this approach is different from our standard model, we do not
explore it further here.

We have also tested a model that includes three phases of gas,
using $\me$ in addition to the two phases in the standard model.
The mass within $\me$ takes into account all the
gas particles that were ejected from the subhalo. These are part of $\mh$ in the
standard two-phase model. For most of the
results quoted below, this model shows a similar agreement to the
HYD, and is thus not discussed in detail. We present the feedback
and cooling efficiencies computed for this model in the Appendix.

\subsection{Star formation}

In Fig.~\ref{fig:fs_tilde} we plot SF efficiencies for the one-phase model,
describing the efficiency of transforming the total amount of gas within a
subhalo into stars. The values of $\fst$ combine the behaviour of $\fs$, $\fc$,
and $\ffd$ from the standard model into a one, compact form (see Eq.~\ref{eq:ode_1phase}).

The values of $\fst$ shown in Fig.~\ref{fig:fs_tilde} are different from what is usually
assumed within one-phase models \citep{Bouche10,Krumholz11,Khochfar11a,Dave11}. For example,
the SF efficiency derived from our HYD changes rapidly as a function of subhalo mass,
while other works assume a constant dependence on subhalo mass.
As was mentioned above, there are other important differences between the model
used here and previous studies. Here the gas mass corresponds to all the gas
inside the subhalo, including both cold and hot components, while previous studies
have used the cold gas only. In addition, previous
models were applied to average main-progenitor histories, while here
the model is evolved through the complicated structure of merger-trees.

\begin{figure}
\centerline{ \hbox{ \epsfig{file=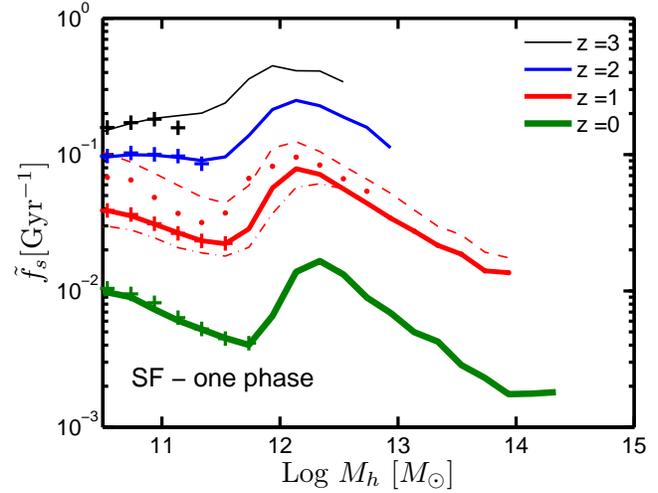,width=9cm} }}
\caption{Star-formation efficiencies within the one-phase model, $\fst$,
as defined in Eq.~\ref{eq:ode_1phase}.
The line-types are the same as in Fig.~\ref{fig:fa}. Plus symbols
are showing the average values of $\fst$ using a sample of subhaloes
with mass lower than $\sim10^{12}\,\msun$ at $\z=0$, along with all their
progenitors.}
  \label{fig:fs_tilde}
\end{figure}

\subsection{Galaxies in the one-phase model}

\begin{figure}
\centerline{ \hbox{ \epsfig{file=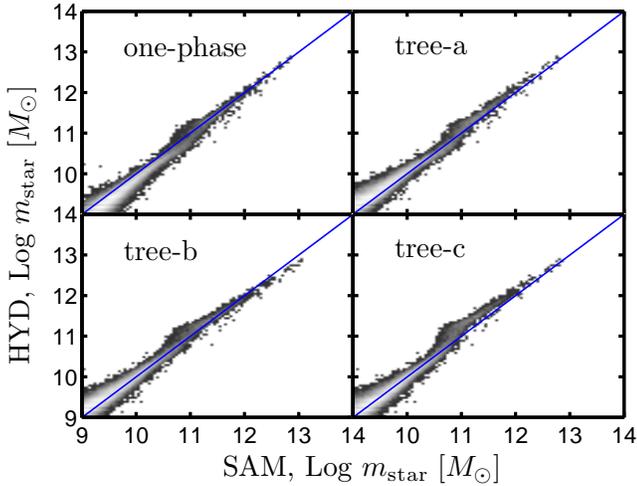,width=9cm} }}
\caption{Comparison of the SAM against the HYD, using deviations for
individual galaxies. The histogram were derived in the same way as in
Fig.~\ref{fig:gal_comp_z0}. Here we plot only stellar masses at $\z=0$.
The upper left panel shows the results of the one-phase model (section
\ref{sec:1phase_eq}). Other panels show a comparison using all galaxies
within the box, using SAM efficiencies that are extracted from the history
of only \emph{one} massive galaxy at $\z=0$. Each panel is based on
efficiencies extracted from a different galaxy, randomly selected.}
\label{fig:gal_comp_test}
\end{figure}

In Fig.~\ref{fig:gal_comp_test} (upper left panel) we compare the
results of the one-phase SAM discussed above, to the original HYD galaxies.
The SAM is based on the efficiency shown in
Fig.~\ref{fig:fs_tilde}. Although the one-phase
model is significantly simpler than our standard model, the results of this model
agree well with the HYD, at the same level as in our standard
model (there is, however, a small systematic deviation of $\sim$0.1 dex
that is seen only in low mass galaxies within the one-phase model). 
We have checked that a similar agreement is obtained at higher
redshifts. To summarize, we find the same accuracy in matching the
SAM against the HYD when using one, two or three gas phases for each
galaxy (the same is true also for the gas components).

The agreement found for the one-phase model proves that the SAM
equations are not unique, and that different models can accurately reproduce
the same HYD. We will discuss the reasons for this
behaviour, and the related implications in section \ref{sec:why_agree}.

\section{More tests}
\label{sec:tests}

\subsection{Efficiencies based on one galaxy}

The efficiencies plotted in Figs.~\ref{fig:fa}-\ref{fig:fd} have
usually a large scatter, when computing the variance over the full
population of galaxies within the box. However, we also show that
once efficiencies are first averaged over all the progenitors of
each $\z=0$ galaxy, the variance decreases significantly. This is
shown as the dotted-dashed lines in the efficiency plots.
As a result, it should be
possible to extract the SAM efficiencies using the history of one
massive galaxy within the HYD.

Fig.~\ref{fig:gal_comp_test} shows results of various different
SAMs. In each model we extracted the SAM efficiencies using
only one random massive galaxy within the HYD, along with all its progenitors.
The results of these models nicely reproduce the population of
the HYD galaxies within the \emph{full} box. This test proves that one
can use a single high-resolution zoom simulation in order to extract the
net result of the physics
of a HYD. Once the efficiencies are known, the SAM can use a large
statistics of merger-trees, based on dark-matter only simulations,
to explore the same physics as was used in the high-resolution HYD.

At stellar masses of $\sim10^{11}\,\msun$ the agreement between
the SAM and HYD is less good. As was discussed in section \ref{sec:comparison},
this might indicate on a larger variation in feedback efficiencies between
subhaloes of the same mass.

\subsection{The dependence of efficiencies on time-steps and on subhalo mass}

The accuracy of the efficiencies discussed here might depend on the
number of output snapshots used to extract information from the HYD.
For example, since the
feedback delay time is 30 Myr after any SF episode, and wind particles
are not allowed to form stars for an additional 15 Myr, the snapshot spacing
should not go below $\sim50$ Myr. To explore the sensitivity of our
method to the time between snapshots, we have tested the same method using half
of the snapshots in our main run. The full analysis described above
was repeated using half of the snapshots, including merger-trees,
computation of efficiencies, and running the SAM. We did not find
any significant changes in the results, and we conclude that our
method is not sensitive to the specific choice of snapshot spacing.
It might be that HYDs which include different physical ingredients
will show more sensitivity in this respect.

In Fig.~\ref{fig:fs_tilde} we plot in symbols the efficiency,
$\fst$, using a range of low mass subhaloes ($\lesssim10^{12}\,\msun$)
within the HYD. We select subhaloes at $\z=0$ and use all their progenitors
for computing the average $\fst$. It is clear that the average $\fst$ for these subhaloes
agrees well with the total average within the box. This point allows
us to use our method in order to increase the dynamical range of
HYDs. Efficiencies for massive subhaloes can in general be extracted
using low resolution simulations. On the other hand, efficiencies
for low mass subhaloes can be derived using a high-resolution
simulation of low mass objects.

\subsection{Why do SAM \& HYD agree?}
\label{sec:why_agree}

In the previous sections we have shown that our SAM agrees
quite well with the HYD, even when comparing individual galaxies.
This fact seems to be surprising. After all, the HYD follows the
dynamics of typically $10^3-10^6$ particles within a galaxy, and the
rates of cooling, SF, and feedback should be affected by many
details within the HYD. For example, metallicity, the abundance of
different elements within the gaseous halo, and the gas density
profile should all affect the cooling rate strongly \citep{Mo10}. It is also
known that the morphology of the galaxy, and the spatial distribution of
cold gas within the disk, should affect the SF rates \citep{McKee07}.
How is it that
the simple SAM, based on average efficiencies per subhalo mass and
time, can reproduce all these dependencies accurately?

The fact that the simple, one-phase model, provides an accurate
match to individual galaxies within the HYD might help us to discover the
reasons for the good agreement. We will thus build our arguments using
the one-phase model, since it is simpler, and includes an analytic
integral solution.

Assume that for a specific galaxy within the HYD, the values
of $\fst$ deviate from the average values as adopted by the SAM. From
Fig.~\ref{fig:fs_tilde}, it looks like deviations between different
merger-trees are negligible. This means that within the full history
of one galaxy (including all its progenitors), the deviations in
$\fst$ with respect to the SAM efficiencies should average to zero.
In order to understand how these
deviations affect $\ms$ we examine the analytic solution presented
in Eqs.~\ref{eq:P} \& \ref{eq:mg_solution}. As seen from these
equations, the mass components of the galaxy depend on integral
values, such as the integral of $\fst$ over time. Since the
deviations in $\fst$ should be averaged out over time, these
deviations do not affect $\ms$ much. This is also true if
the model is not ideal in the sense that $\fst$ depends on $\mg$.

The fact that our efficiencies are computed using average values per subhalo
mass and time is the other key ingredient for the good match between the
SAM and the HYD. Such a method guarantees that on average, the mass
of each baryonic component within the SAM will agree with the HYD.
Moreover, the scatter between the models should be small because various
random processes cancel each other:
the SF history of one galaxy is built from
different episodes in its history, and the stellar mass is a sum over
all these episodes. If each SF event within the HYD includes some random error
(with respect to the SAM), the sum of all should have a much smaller
scatter. This is the reason why the deviations in SF rates
between the HYD and SAM are of the order of 0.5 dex,
while the total stellar masses agree much better, to a level of 0.1 dex.

\section{Summary and discussion}
\label{sec:discuss}

In this work we have developed a method to post-process a
hydrodynamical simulation of galaxy formation (HYD), and to extract the
simple baryonic laws that shape the evolution of galaxies within it.
By using the same laws within a semi-analytic model (SAM), we
confirmed that the resulting galaxies in both models remain almost unchanged.

We have used a state-of-the-art HYD, taken from the OWLS project \citep{Schaye10}.
This simulation includes radiative cooling, a galactic wind model for feedback, and a SF
recipe that mimics the observed Kennicutt-Schmidt law. On the other
hand, we have used a simple SAM, based on the approach presented in
\citet{Neistein10}. In this SAM, the processes of accretion,
cooling, SF, and feedback, are modeled using efficiencies that
depend only on the host subhalo mass and redshift.
Although previous studies did not find a good agreement
between HYDs and SAMs, our simple SAM can reproduce the results of the
advanced HYD at the level of 0.1 dex, when comparing individual galaxies
(see Fig.~\ref{fig:gal_comp_z0}).
The same level of accuracy is achieved for all redshifts and for various
mass components (stars and gas, although
the cold gas component shows deviations of $\sim0.2$ dex). Statistical
properties of galaxies, like the stellar mass function, are
accurately reproduced by the SAM as well (see Fig.~\ref{fig:mass_funs}).

We claimed that the main reason for the agreement between the SAM and
the HYD is the nature of the differential equations that govern galaxy
evolution: Using a simple one-phase model, that can be solved
analytically, we show that the mass components depend on the
integrals of efficiencies over time, so temporal variations
cannot be seen in the total stellar mass
of galaxies. On the other hand, instantaneous properties like SF
rates, reveal much larger deviations between the SAM and the HYD,
at the level of 0.5 dex. The fact that our recipes are matched for
the same subhalo mass, implies that objects from the SAM and
HYD should agree on average, since the common reference between these
models is the merger-tree, based on the subhalo mass.

We have found that satellite galaxies within the HYD experience stripping of hot
gas, on a time scale of $\sim3$ Gyr. Although we have used only one time-scale for
stripping, we found that a more accurate description would demand a
time-scale that depends on the satellite mass, and on the infall redshift.
In addition, satellite galaxies have different efficiencies of cooling and
feedback than central galaxies, a fact that was neglected in
this work. Due to the above simplifications made in our model, the
agreement between satellite galaxies in the HYD and SAM reaches a
value of 0.2 dex at low redshift.

The method presented here is very robust and should work for any
hydrodynamical simulation. We have shown that it even works if we
assume only one phase of gas within subhaloes. In this highly simplified
SAM there are only two processes: smooth accretion, and SF.
Nonetheless, the SAM is able to reproduce the same galaxies as in
the HYD with the same accuracy as a SAM with more phases of gas 
(see Fig.~\ref{fig:gal_comp_test}).
This test proves that our method is insensitive to the specific
way the SAM equations are written, and that there are many possible
models to describe the same simulation.

We have shown that the variance in the baryonic efficiencies is very
small, once efficiencies are first averaged over
the progenitors of each $\z=0$ galaxy.
We have also extracted the SAM efficiencies from the progenitors of
one random massive galaxy at $\z=0$. These efficiencies are then
able to reproduce the full population of galaxies within the HYD.
Consequently, our method can separate between baryonic processes
and cosmic variance, and can be used to interpret one zoomed
hydrodynamical simulation. In a future work, we plan to use the method developed here
in order to scan combinations of various implementations of feedback and SF within hydrodynamical
simulations.

We have investigated
the efficiencies of the baryonic processes within 
the HYD, showing that this specific
simulation deviates from the values adopted by standard SAMs.
We have shown that smooth accretion does not always
follow dark-matter, and is less efficient for low-mass haloes \citep[see also][]{vdVoort11}.
All the fresh gas falling into subhaloes is either hot or dilute,
and cannot form stars. For a given subhalo mass, cooling time-scales are roughly
proportional to the
cosmic time, with somewhat shorter time-scales at high-$\z$. On
the other hand, the efficiency of SF does not change significantly
with cosmic time and does not decrease strongly for high-mass
haloes. Lastly, the HYD used here does not show noticeable contributions
from merger-induced star formation bursts.
We hope that these findings can help to bridge the gap
between SAMs and HYDs, and will be useful for interpreting
various existing models.


\section*{Acknowledgments}

We thank the referee for a detailed and constructive report that helped us to 
improve the paper.
We thank Volker Springel for making his code \textsc{gadget}
publicly available, and for letting us use the \textsc{subfind} code.
EN and SK acknowledge funding by the DFG via grant KH-254/2-1.
CDV is supported by Marie Curie Reintegration Grant FP7-RG-256573.
We acknowledge useful discussions with Till Sawala.

\bibliographystyle{mn2e}
\bibliography{ref_list}

\appendix
\section{More efficiencies}

\begin{figure}
\centerline{ \hbox{ \epsfig{file=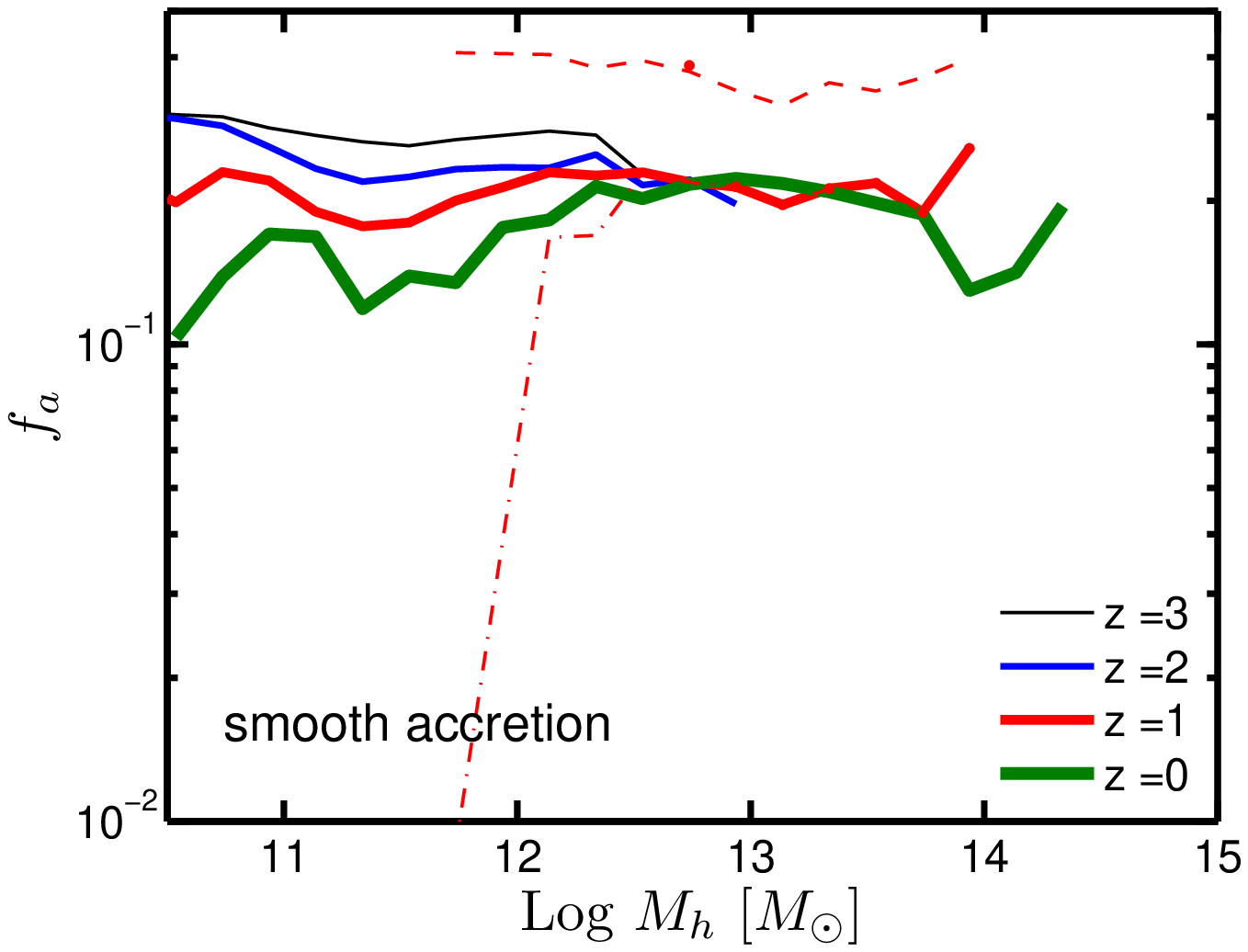,width=9cm} }}
\caption{The smooth accretion rate of baryons as derived from the hydrodynamical
simulation, using a model in which stripping of hot gas follows the dark-matter stripping.
The definitions of line-types are the same as in Fig.~\ref{fig:fa}.}
  \label{fig:fa_strip}
\end{figure}

In this section we plot additional efficiency values that were extracted
from the HYD. In Fig.~\ref{fig:fa_strip} we plot the smooth
accretion, when stripping is allowed for central galaxies. These
accretion rates are very different from our standard model
(Fig.~\ref{fig:fa}), proving that stripping occurs often in low
mass haloes.

\begin{figure}
\centerline{ \hbox{ \epsfig{file=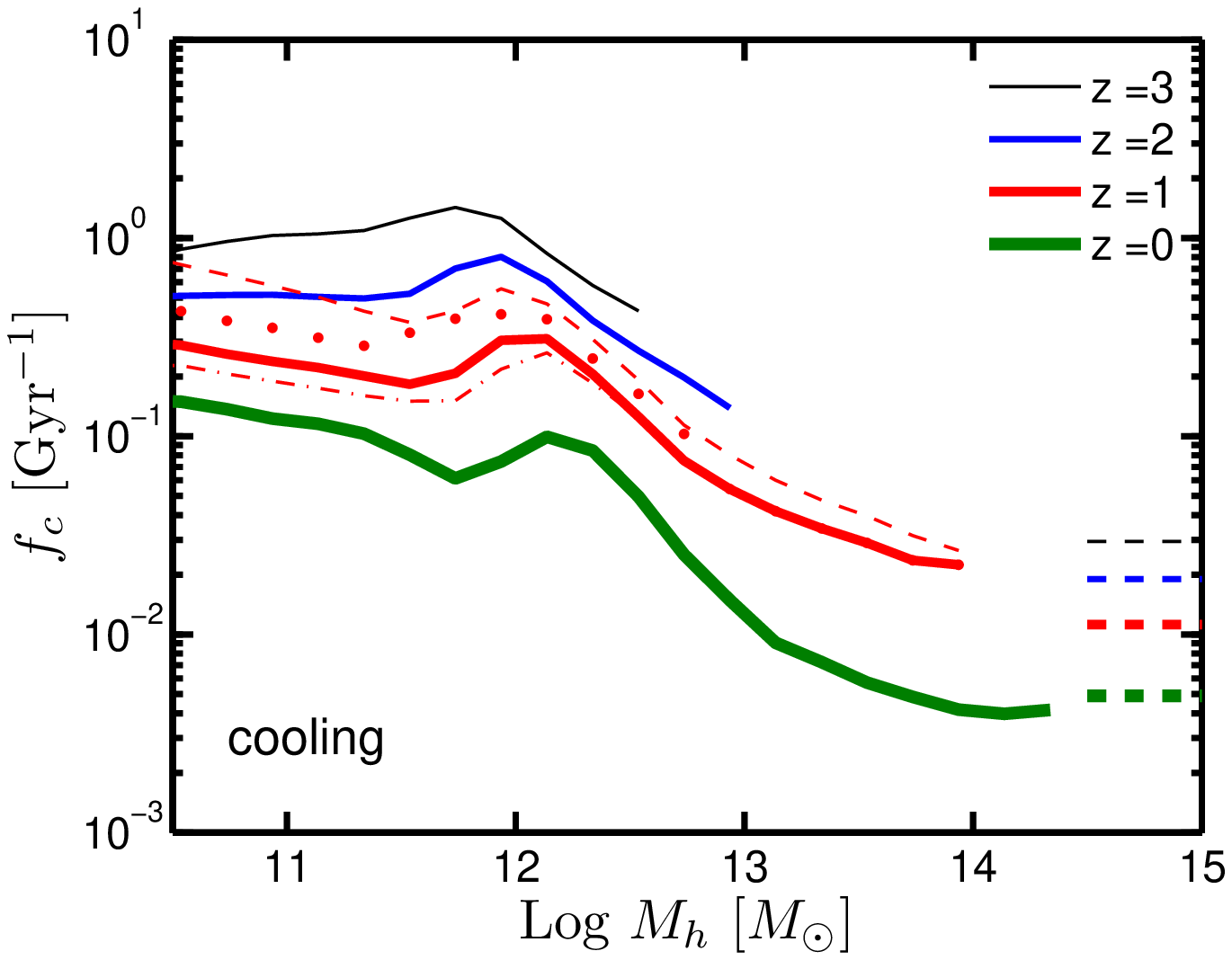,width=9cm} }}
\centerline{ \hbox{ \epsfig{file=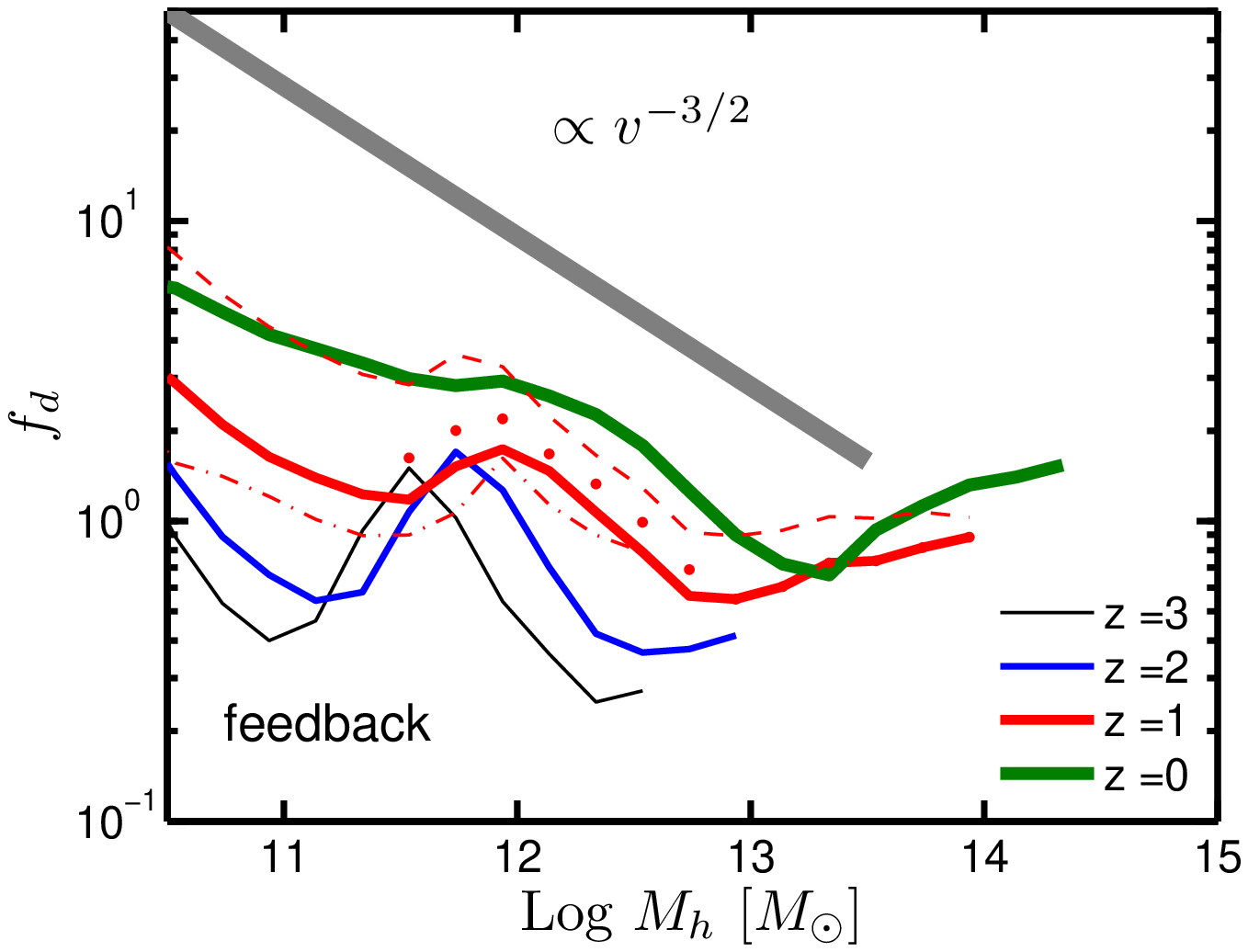,width=9cm} }}
\caption{The cooling and feedback efficiencies within the HYD, using a separate ejection phase.
The definitions of line-types are the same as in Fig.~\ref{fig:fa}.}
  \label{fig:fcd_ejc}
\end{figure}

In Fig.~\ref{fig:fcd_ejc} we show the cooling and feedback
efficiencies when we use an additional gas phase, $\me$ to
describe the ejected gas. Here $\fc$ describes the transition from
$\mh$ to $\mc$ only, and is normalized by $\mh$. Feedback is defined
as the total heated gas, including all gas particles that started within
$\mc$ and ended within $\mh$ or $\me$.

\begin{figure}
\centerline{ \hbox{ \epsfig{file=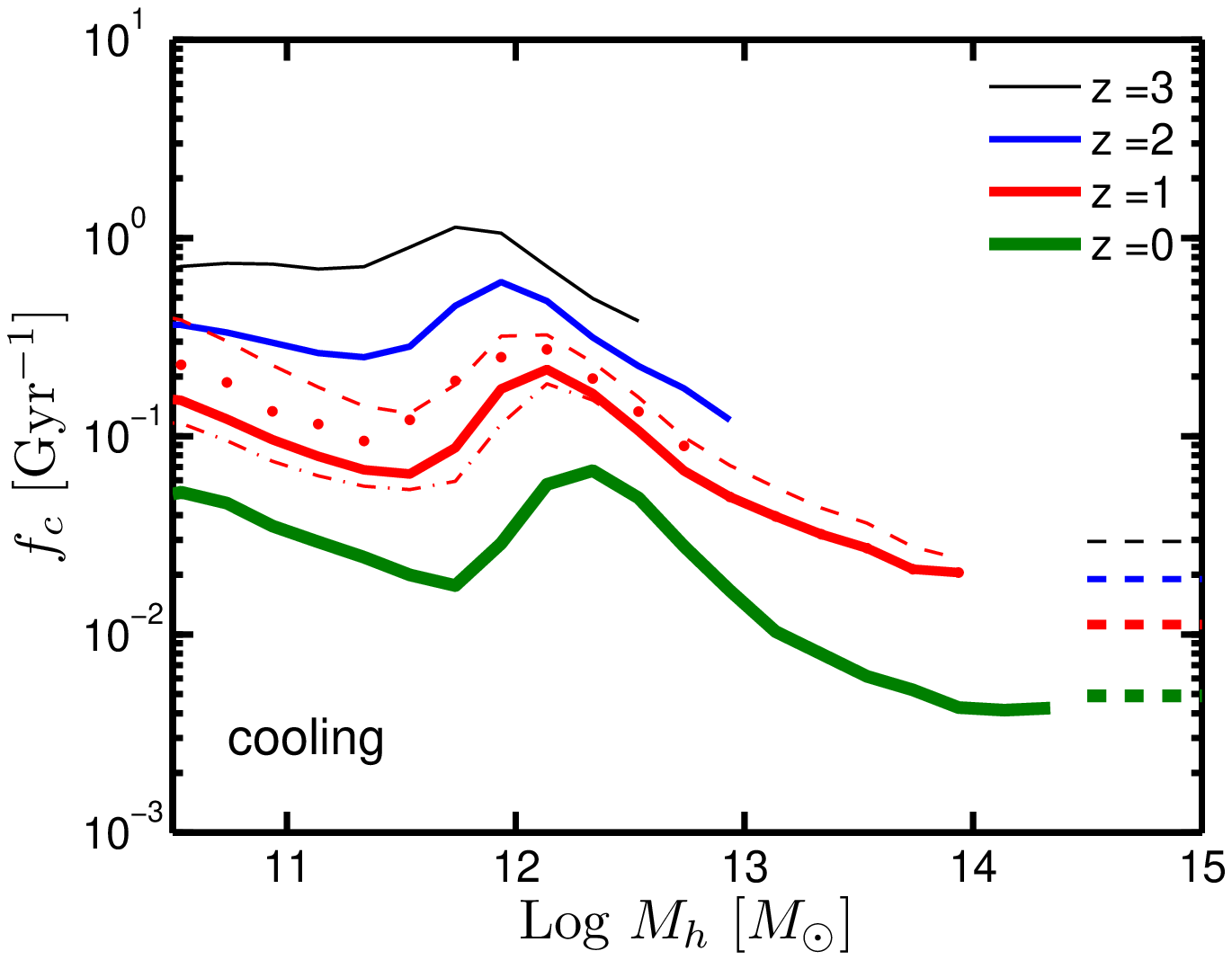,width=9cm} }}
\centerline{ \hbox{ \epsfig{file=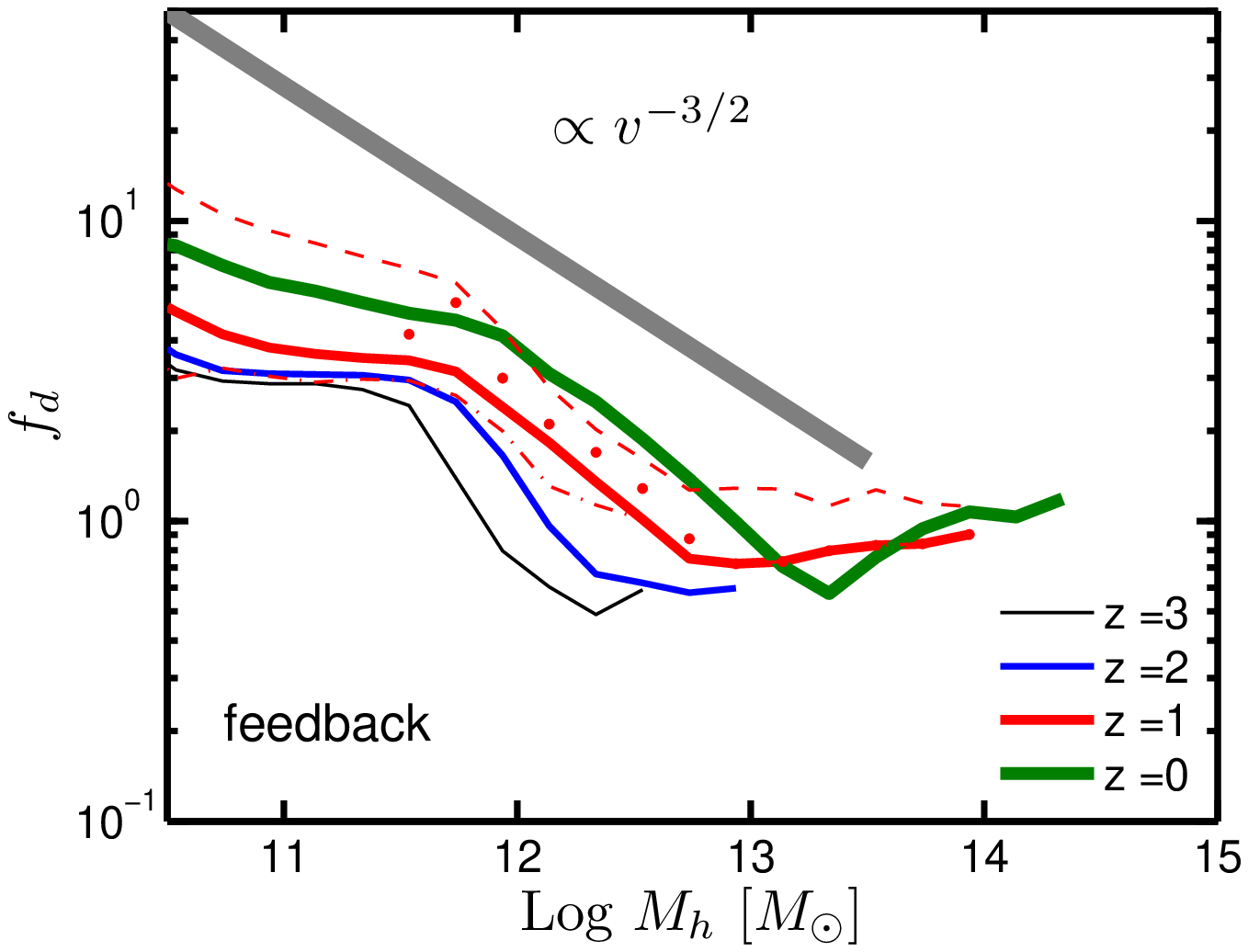,width=9cm} }}
\caption{The cooling and feedback efficiencies using the proper mass for each particle
within the HYD. The definitions of line-types are the same as in Fig.~\ref{fig:fa}.}
  \label{fig:fcd_proper}
\end{figure}

\begin{figure}
\centerline{ \hbox{ \epsfig{file=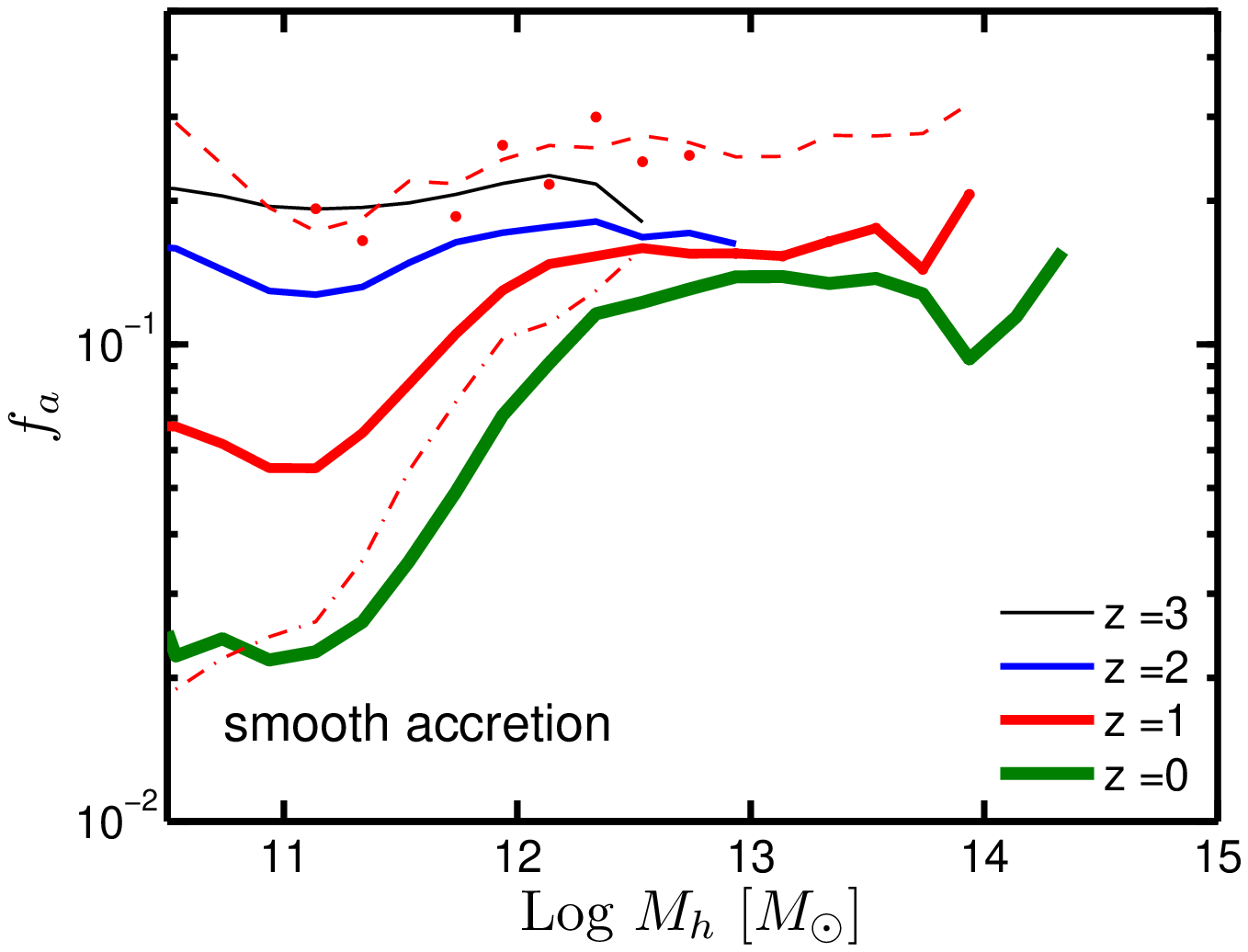,width=9cm} }}
\centerline{ \hbox{ \epsfig{file=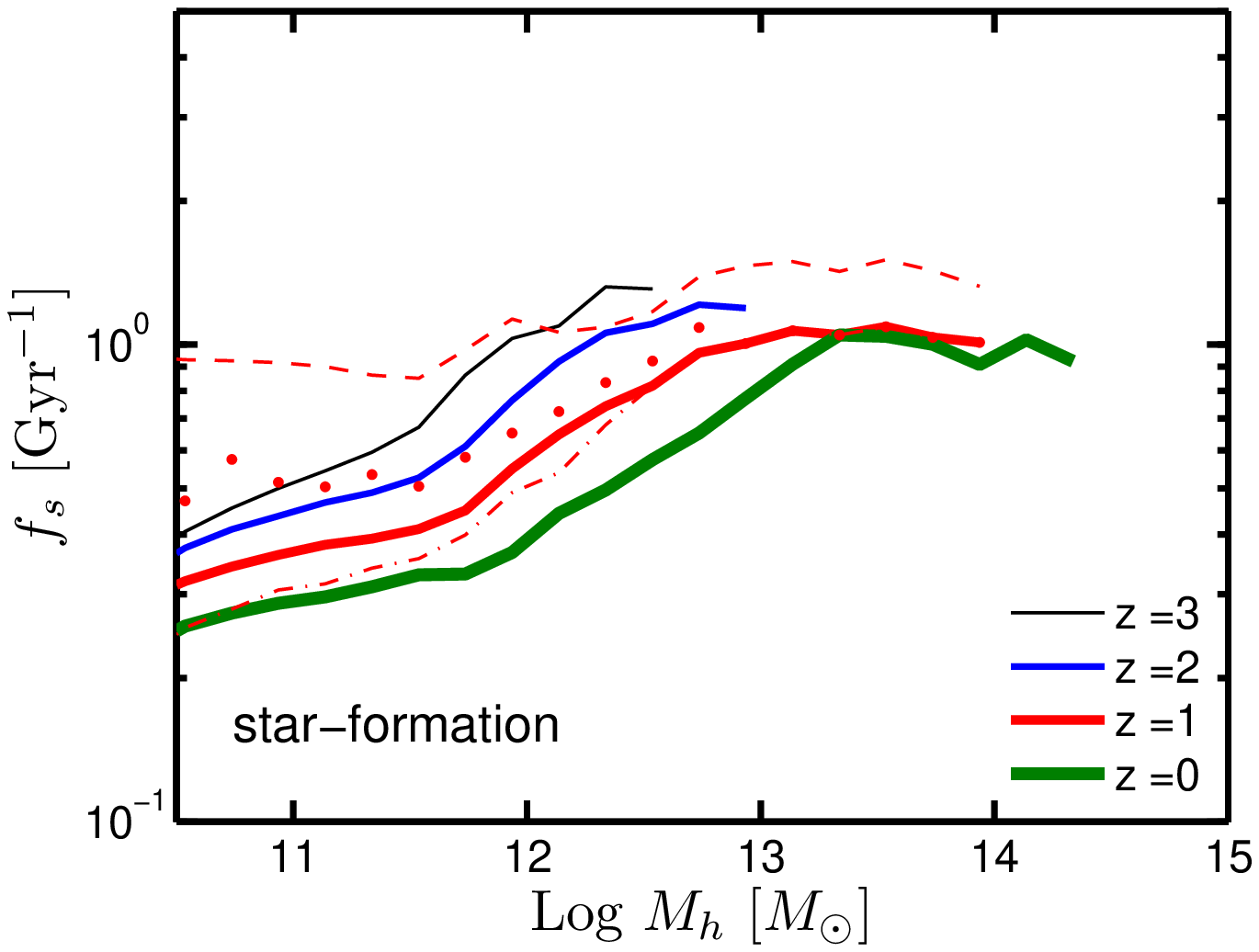,width=9cm} }}
\caption{The accretion and SF efficiencies using the proper mass for each particle
within the HYD. The definitions of line-types are the same as in Fig.~\ref{fig:fa}.}
  \label{fig:fas_proper}
\end{figure}

The efficiencies plotted in Figs.~\ref{fig:fa}-\ref{fig:fd}
use a fixed particle mass, and do not account for
mass loss due to SN and
stellar winds. Although these efficiencies were treated
self-consistently in the main body of this work, they might deviate
from the true efficiencies, computed with the proper particle mass.
In Figs.~\ref{fig:fcd_proper} \& \ref{fig:fas_proper} we plot the
true efficiencies. These are very similar to the values used before,
but show some small differences in the overall normalization.

\label{lastpage}

\end{document}